\documentclass[journal=langd5,manuscript=article]{achemso}

\usepackage{chemformula} 
\usepackage[T1]{fontenc} 
\usepackage[utf8]{inputenc}

\usepackage{xr}

\author{Anne C. Nickel}
\affiliation[RWTH Aachen University]
{Institute of Physical Chemistry,  RWTH Aachen University, 52056 Aachen, Germany, European Union}
\author{Timon Kratzenberg}
\affiliation[RWTH Aachen University]
{Institute of Physical Chemistry,  RWTH Aachen University, 52056 Aachen, Germany, European Union}
\author{Steffen Bochenek}
\affiliation[RWTH Aachen University]
{Institute of Physical Chemistry,  RWTH Aachen University, 52056 Aachen, Germany, European Union}
\author{Maximilian M. Schmidt}
\affiliation[RWTH Aachen University]
{Institute of Physical Chemistry,  RWTH Aachen University, 52056 Aachen, Germany, European Union}
\author{Andrey A. Rudov}
\affiliation[Moscow State University]
{Physics Department, Lomonosov Moscow State University, Moscow 119991, Russian Federation}
\alsoaffiliation[DWI]
{DWI - Leibniz-Institute for Interactive Materials, 52056 Aachen, Germany, European Union}
\author{Andreas Falkenstein}
\affiliation[RWTH Aachen University]
{Institute of Physical Chemistry,  RWTH Aachen University, 52056 Aachen, Germany, European Union}
\author{Igor I. Potemkin}
\affiliation[Moscow State University]
{Physics Department, Lomonosov Moscow State University, Moscow 119991, Russian Federation}
\alsoaffiliation[DWI]
{DWI - Leibniz-Institute for Interactive Materials, 52056 Aachen, Germany, European Union}
\alsoaffiliation[Ural]
{National Research South Ural State University, Chelyabinsk 454080, Russian Federation}
\author{Jérôme J. Crassous}
\affiliation[RWTH Aachen University]
{Institute of Physical Chemistry,  RWTH Aachen University, 52056 Aachen, Germany, European Union}
\author{Walter Richtering}
\email{richtering@rwth-aachen.de}
\affiliation[RWTH Aachen University]
{Institute of Physical Chemistry,  RWTH Aachen University, 52056 Aachen, Germany, European Union}

\title[Anisotropic Microgels show their Soft Side]
 {Anisotropic Microgels show their Soft Side}

\begin{document}

\begin{abstract}
Anisotropic, submicrometer sized particles are versatile systems providing interesting features in creating ordering in 2-dimensional systems. Combining hard ellipsoids with a soft shell further enhances the opportunities to trigger and control order and alignment. In this work, we report a rich 2D phase behavior and show how softness affects the ordering of anisotropic particles at fluid oil-water interfaces. Three different core-shell systems were synthesized such that they have the same elliptical hematite-silica core but differ with respect to thickness and stiffness of the soft microgel shell. Compression isotherms, the shape of individual core-shell microgel as well as their 2D order at a decane-water interface are investigated by means of the Langmuir-Blodgett technique combined with ex-situ Atomic Force Microscopy (AFM) imaging, as well as by  Disspiative Particle Dynamics (DPD) simulations. We show how softness, size and anisotropy of the microgel shell affect the side-to-side- vs. tip-to-tip ordering of anisotropic hybrid microgels as well as the alignment with respect to the direction of compression in the Langmuir trough. A large and soft microgel shell leads to an ordered structure with a tip-to-tip alignment directed perpendicular to the direction of compression. In contrast, a thin and harder microgel shell leads to side-to-side ordering orientated parallel to the compression direction. In addition, the thin and harder microgel shell induces clustering of the microgels in the dilute state indicating the presence of strong capillary interactions. Our findings highlight the relevance of softness for the complex ordering of anisotropic hybrid microgels at interfaces.  

\end{abstract}

\section{Introduction}
\begin{figure}[ht]
\vspace{0.5cm}
\centering
\includegraphics[width=0.5\linewidth, trim={0cm 0cm 0cm 0cm},clip]{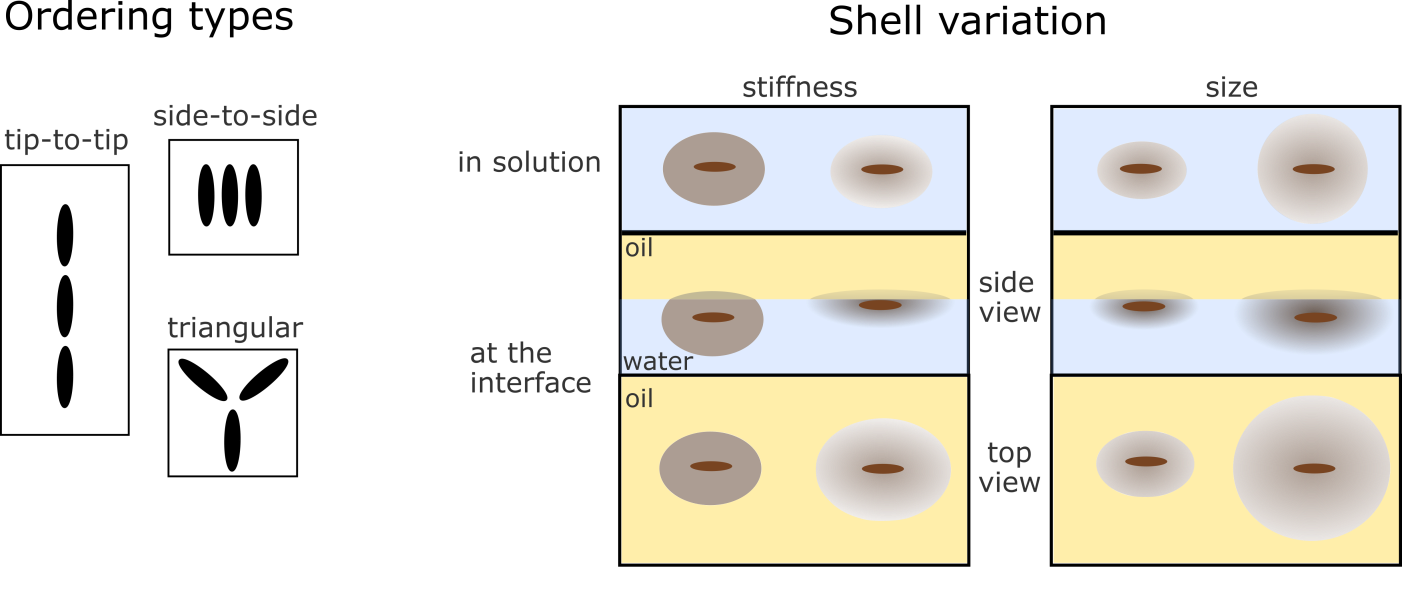}
	\caption{Tip-to-tip, side-to-side and triangular are possible ordering types for hard ellipsoids (left). By combining a microgel shell with the elliptical core, the softness and the size of the hybrid microgels can be adjusted (right).}
	\label{MainQuestion}
\end{figure}

Particles in the submicrometer size are used as model systems to exploit physical properties for example in biological applications or as emulsion stabilizers and are therefore widely examined in literature.\cite{Wiese2013,Karg:2019kf,Dirksen2020,Rey2020} Especially their ordering in a 2-dimensional state is interesting as the surface coverage is important for  applications.\cite{Honold2015,Xia2013} As most biological systems are soft, microgels, cross-linked polymers swollen in a good solvent\cite{Karg:2019kf}, are an ideal model system. Compared to hard nanoparticles, the temperature-responsiveness and softness of microgels can be readily tuned by the monomer composition.\cite{KRATZ20016631,Halperin2015} The reaction kinetics of the monomer \textit{N}-isopropylacrylamide (NIPAm) and the cross-linker \textit{N},\textit{N'}-methylene-bisacrylamide (BIS) lead to microgels with a higher cross-linked core surrounded by a fuzzy corona.\cite{Stieger2004} As biological systems like bacteria are not only soft but additionally anisotropic, anisotropic microgels are an interesting model system to study.\\
The shape effect of the ordering of bacteria\cite{Dell2018} and the misfolding of proteins\cite{Merlini2003} are some examples. Hence, understanding the ordering phenomena of differently shaped model systems is of great interest in today's research. Additionally, the physical properties become more complicated when moving from spherical nanoparticles to anisotropically shaped nanoparticles.\cite{Boeker2007,Grzelczak2010} As a result, the possible applications widen. For example, anisotropic particles and especially their shape-dependent ordering are important for optics\cite{Ganesan2017} and shape-dependent emulsion stabilizers\cite{Madivala2009}.\\
\noindent Anisotropic particles exhibit different orientations towards their nearest neighbors: tip-to-tip (t-t)\cite{Loudet2005}, side-to-side (s-s)\cite{Loudet2005,Basavaraj2009,Pietra2012} or triangular\cite{Basavaraj2009}. Which ordering occurs can be explained by capillary interactions inducing self-assembly. The anisotropic particles rotate leading to a t-t contact for far-field capillary interactions. In contrast to that, the specific shape of the particles becomes relevant for near-field capillary interactions. Ellipsoidal particles minimize the capillary energy by aligning s-s whereas cylindrical particles prefer t-t orientation but can also align in s-s.\cite{Botto2012} This is caused by a curved contact line to fulfill Young's equation.\cite{Lehle2008} The contact line for cylinders rises at the flat ends and is lowered at the sides. Instead, the contact line for ellipsoids rises at the sides and is lowered at the tips.\cite{Lehle2008, Botto2012, Li2019} This complex connection between shape and self-assembly is additionally influenced by the polydispersity of the particles. For example, Loudet et al. showed that polydisperse ellipsoids assemble in arrow-like structures.\cite{Loudet2009} In conclusion, the ordering behavior and phase transitions of anisotropic particles depend on multiple factors such as aspect ratio\cite{Basavaraj2009}, total size, shape,\cite{Botto2012} and hardness\cite{Honda2019}. \\
\noindent To maintain the benefits of anisotropic structures and expand these with the advantages of soft materials, a combination via a core-shell structure is possible. One way to obtain a soft shell is by synthesizing cross-linked polymers onto an anisotropic hard core. These anisotropic core-shell microgels are obtained by seed and feed precipitation polymerization onto elliptical hematite cores.\cite{Crassous2014,Nickel2019}\\
Microgels are highly interfacial active and lower the interfacial tension, thereby they are deformed and spread at the interface.\cite{Geisel2012,Mihut2013,Pinaud2014,Camerin2019} Due to their small size, microgels are difficult to visualize at the decane-water interface. Hence, spherical microgels and their ordering and monolayer formation at decane-water interfaces are commonly investigated with the Langmuir-Blodgett technique and simultaneous deposition of the monolayer onto a solid substrate. These substrates are used to investigate the monolayers ex situ with atomic force microscopy (AFM). A combination of the Langmuir-Blodgett technique and AFM enables to probe the structure of the microgel monolayer at different compression states.\cite{Geisel2014,Picard2017,Bochenek2019} \\
Spherical particles and additionally spherical microgels and the different regions appearing while compressing the monolayer are studied intensely.\cite{Aveyard2000,Rey2016,Picard2017,Bochenek2019,Schmidt:2020kv,Ciarella:2021a} Due to their core-corona structure, spherical microgels exhibit five regions during their compression\cite{Pinaud2014,Rey2016,Bochenek2019} while for hard spherical particles three regions are observed in the compression isotherm\cite{Aveyard2000}. The steep increase of the surface pressure in the second region is a result of the repulsion between the hard spherical particles.\cite{Aveyard2000} Furthermore, the combination of a hard spherical particle with a microgel shell leads to a similar compression isotherm with five regions.\cite{Geisel2015}\\
The ordering phenomena of anisotropic particles especially anisotropic microgels at interfaces is not studied in such detail. A recent study by Honda et al. showed that the combination of a soft microgel shell onto a hard ellipsoidal core has a major influence on their self-assembly when compared to the self-assembly of the hard core without the soft shell. The cores themselves lose their s-s ordering while the core-shell microgels keep their s-s ordering during the evaporation of an aqueous droplet, hence, a combination of an increase in concentration and a progress in time.\cite{Honda2019} Nevertheless, the compression isotherms of hybrid core-shell microgels and the effect on the ordering of different microgel shells have not been investigated yet. 
In addition, the phase behavior of anisotropic particles is influenced by their aspect ratio, size, wetting angle and contact line.\cite{Basavaraj2006,Ballard2015,Luo2019} For example, the shape of the compression isotherms differentiates when ellipsoids with different aspect ratios are compared.\cite{Basavaraj2006} Similar to the spherical case an even more complex behavior for compression isotherms can be envisioned when a microgel shell is added to an anisotropic core.\\
Recent studies have focused either on the ordering of soft spherically shaped materials, e.g. spherical microgels\cite{Bochenek2019}, or hard anisotropically shaped materials\cite{Anjali2019} in 2-dimensional crowded environments. However, no one has so far investigated how the ordering of anisotropic hybrid core-shell microgels in a crowded environment is influenced by their shell characteristics. Like for hard ellipsoidal particles, the three different ordering types t-t, s-s or triangular are possible (see Figure~\ref{MainQuestion} left). The variation of the shell by adjusting the microgel stiffness and size could lead to different hybrid microgels with different ordering behavior. Figure~\ref{MainQuestion} (right) illustrates that due to the different spreading of the microgel shells, soft or thick microgel shells occupy larger area at the interface compared to hard or thin microgel shells. Such hybrid microgels allow obtaining fundamental understanding of what softness does to the anisotropy of anisometric systems and how it influences their interfacial assembly.

This study focuses on anisotropically shaped hybrid core-shell microgels compressed at a decane-water interface. Three core-shell microgel systems are investigated with a Langmuir-Blodgett trough experimentally and complementary with dissipative dynamics simulations. The systems consist of the same elliptical hard hematite-silica core and differ within the softness and thickness of their microgel shell. Experimentally the monolayers are deposited on silica substrates to investigate their ordering and the structure of the microgels ex situ with atomic force microscopy (AFM). \\
With the combination of experiments and simulations, we reveal that the ordering and, as a result, the self-assembly of anisotropic core-shell microgels at an interface depends mainly on the thickness and the softness of the microgel shell. By using different shells, we can fine tune capillary and exclude volume interactions, which are controlled by aspect ratio, i.e., shape of the spread shell, and the self-assembly of the microgels, leading to either preferred s-s, t-t ordering or to no order at all. Especially the simulations allow us to follow the change of the anisotropy of the microgels, the redistribution of polymeric density and to track the relative position of the solid cores and their rotation at the interface upon compression. Experimentally, we show that microgels align differently due to the compression exerted by the Langmuir-Blodgett trough barriers. For a thick microgel shell, the microgel cores align perpendicular to the direction of compression and show t-t ordering. For a thin microgel shell, s-s clustering occurs and the microgel cores align parallel to the direction of the compression.\\

\section{Experimental Section}
\textbf{Materials} All materials were used as purchased. The cross-linker BIS (\textit{N},\textit{N'}-methylene-bisacrylamide) was purchased from Sigma-Aldrich and the main monomer NIPAm (\textit{N}-isopropylacrylamide) from Acro organics. The initiator KPS (potassium peroxydisulfate) was bought from Merck. The co-monomer BAC (\textit{N},\textit{N}’-bis(acryloyl)cystamine) was purchased from Alfa Aesar. For the Langmuir-Blodgett trough experiments decane (Merck) , ultrapure water (Astacus$^{2}$, membraPure GmbH, Germany) and propan-2-ol (Merck) were used. The decane was cleaned by filtering it three times over a column of basic Al$_{2}$O$_{3}$ (Merck). The third filtering was done just before the measurement. \\
\textbf{Synthesis} The synthesis procedure of the microgels CS-356 and CS-254 are described elsewhere in more detail.\cite{Nickel2019} The synthesis of the third microgel (CS-165) was adjusted by replacing the dye methacryloxyethyl thiocarbamoyl rhodamine B (MRB) with the co-monomer BAC. The molar mass composition of the cross-linker BIS was kept constant and the amount of NIPAm was reduced to add 0.9~mol\% of BAC. As BAC acts as an additional cross-linker, the cross-linking density increases to 5.62~mol\% from approximately 4.65~mol\%. The composition with respect to monomer, co-monomer and cross-linker is provided in Table~\ref{table_molar}.

\begin{table}[H]
     \centering
    \caption{Molar mass percentage of main monomer, co-monomer and cross-linker used for the synthesis of CS-165, CS-254 and CS-356.}
     \begin{tabular}{cccc}
       \hline
 & \textit{CS-165} & \textit{CS-254} & \textit{CS-356}\\
       \hline
NIPAm [mol\%] & 94.37 & 95.34 & 95.28\\
BIS [mol\%] & 4.72 & 4.63 & 4.69 \\ 
MRB [mol\%] & 0 & 0.02 & 0.02 \\
BAC [mol\%] & 0.9 & 0 & 0 \\
       \hline
     \end{tabular}
     \label{table_molar}
     \end{table}

\begin{figure}[ht]
\centering
    \includegraphics[width=0.65\linewidth, trim={0cm 0cm 0cm 0cm},clip]{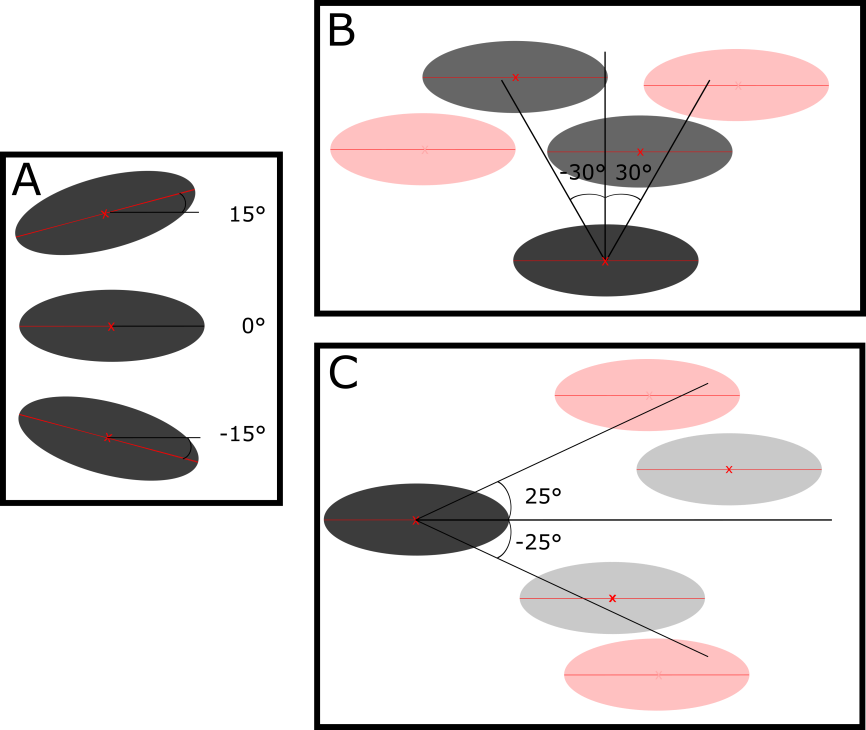}	
	\caption{Sketch of the classification for s-s and t-t ordering in this study. The neighboring microgel has to have an angle difference of not more than $\pm$~15~°to the original microgel (A). The definition of s-s (B) and t-t (C) order is based on the position of the neighboring microgel compared to the orientation of the original microgel.}
	\label{ST_Ordering}
\end{figure}

To synthesize CS-165, 584.4~mg of NIPAm, 39.6~mg of BIS and 14.3~mg of BAC were dissolved in 280~mL of filtered bidistilled water. 3~mL of an ethanol solution with 1.4~wt\% of the same ellipsoidal functionalized hematite-silica cores used in Nickel et al.\cite{Nickel2019} was added into the three-neck flask after sonification for 30~minutes to reduce the amount of aggregated within the core solution. The mixture was heated with an external oil-bath to 60~°C and degased with nitrogen. The initiator solution consisted of 41.0~mg KPS and 10~mL filtered bidestilled water and was degased in the syringe. After heating the reaction solution to 60~°C for 45~Minutes, the oil-bath was set to 80~°C. After another 15~Minutes, the polymerisation was started by adding the initiator solution dropwise. The reaction was stopped after 4~h by removing the oil-bath and the nitrogen flow. The microgels were purified with three times centrifugation at 5000~rpm for 60~Minutes and redispersing in filtered bidestilled water. The cleaned microgel solution was lyophilizates.\\
\textbf{Langmuir-Blodgett technique} The isotherms and deposition were conducted simultaneously at a decane-water interface in a customized liquid-liquid Langmuir-Blodgett trough (KSV NIMA, Biolin Scientific Oy, Finland) equipped with two movable barriers. The trough has an area of around 402~cm$^{2}$ when the barriers are fully open and is made of poly(oxymethylene) glycol. To carry out the experiments at 20~°C, the trough was connected to an external water bath. As the microgels are too small to be visualized in-situ, the monolayer was transferred simultaneously to the compression to a silica substrate (6~cm~$\times$~1~cm) to visualize the monolayers ex-situ.\\
For each measurement, the trough and the barriers were cleaned with ethanol and milli-q water. For the subphase milli-q water was used and the interface was cleaned with a suction pump. To measure the surface pressure, a platinum Wilhelmy plate was used and placed parallel to the barriers in the middle of the trough. The plate was lowered to the surface and then 200~mL of decane were added to the trough to obtain a clean water-decane interface.\\
\textbf{Atomic force microscopy} To visualize the monolayers deposited on the silica substrates, dry-state AFM was used. The experiments were carried out in tapping mode using a Dimension Icon AFM with a closed loop (Veeco Instruments Inc., USA, Software: NanoScope 9.4, Bruker Co., USA). The tips used for imaging were OTESPA tips with a resonance frequency of 300~kHz, a nominal spring constant of 26~N$\cdot$m$^{-1}$ of the cantilever and a nominal tip radius of <7~nm (NanoAndMore GmbH, Germany). The images were taken with the programmed move function to scan the silica substrates. On each substrate, six rows along the gradient of microgel concentration were captured with a step-width of 1000~µm. Each image is 7.5~µm~$\times$~7.5~µm large with a resolution of 512~pixels~$\times$~512~pixels.\\
\textbf{Image analysis} For the quantitative analysis of the AFM images, NanoScope Analysis 1.9 was used to remove the tilt by leveling the images. Zero height of the images was fixed to the minimum values. The processed AFM images were analysed with a modified version of the MATLAB routine of Bochenek et al.\cite{Bochenek2019} Instead of employing the particle tracking code by Crocker and Grier,\cite{Crocker1996} elliptical microgels were located employing the MATLAB function regionprops. Before the microgels could be localized, the images were filtered by applying a band-pass filter and a zero crossing edge detection. Regionprops was used to determined the position, orientation, dimensions, eccentricity and mean brightness of the microgels. To enhance the accuracy, the results were refined by user-defined thresholds regarding the dimensions, eccentricity and mean brightness of the localized microgels. The nearest neighbor connections were obtained through Delaunay triangulation and the Voronoi tessellation. When determining the ordering types, it is necessary to only consider microgels within one cluster. Therefore, a threshold for the nearest neighbor distance was implemented, removing Delaunay edges between different microgel cluster. From this information, our definition of s-s and t-t was applied to obtain information about the percent of both ordering types: When looking at one microgel, each neighboring microgel according to the Voronoi tessellation is checked for the difference in the alignment. The script accepts only the neighboring microgels with a difference in alignment of $\pm$~15° to the alignment of the middle microgels and investigates these neighboring microgels further (Figure~\ref{ST_Ordering}~A). For a randomly ordered sample this would be 16.67~\% (30°/180°) of the neighboring microgels. Hence, this value is used as reference for isotropic ordering behavior. Each neighboring microgel fulfilling the first criteria is checked for s-s or t-t ordering. To count for s-s ordering, the center of the neighboring microgel has to lay in an angle range of $\pm$~30° of the perpendicular alignment of the microgel (Figure~\ref{ST_Ordering}~B marked in grey). For t-t ordering, the neighboring microgel must be located with it's center at an angle range of $\pm$~25° of the alignment of the microgel (Figure~\ref{ST_Ordering}~C marked in light grey). The values were chosen after checking a couple of different values for images at high concentration with a small nearest neighbor distance and for images at low concentration with a larger nearest neighbor distance. Larger angles overestimated the amount of ordered microgels while smaller angles lacked the assigning of ordered microgels with low distances. As the script uses angles and the center of the microgels only, the probability to detect ordering increases for microgels with a larger nearest neighbor distance as the area increases with increasing distance. Hence, the used values assured to detect the s-s ordering of close microgels in a reasonable amount, but to not over-interpret the ordering for larger distances between the microgels. That different angles are chosen for the different ordering types depends on the fact that we use the center of the microgel to check for the ordering. As a result of the anisotropic shape of the microgels, the center of the microgel is less probable located within the area enclosed by the angles at the sides compared to the tips if the microgels are orientated similarly (see Figure~\ref{ST_Ordering}). The angle used for s-s ordering is slightly larger to balance this effect. If at least one of the neighboring microgels accounts for s-s or t-t ordering, the microgel is accounted for the respective ordering. Consequently, each microgel can be accounted for no ordering, s-s ordering, t-t ordering or both ordering types depending on the neighboring microgels.

\textbf{Computer simulations}\\
\textbf{Method} Dissipative particle dynamics (DPD) simulations\cite{Hoogerbrugge1992,Espanol1995,Warren1997} were used to study the behavior of elliptic hybrid core-shell pNIPAm based microgels at the decan-water interface. The method consists of reducing the complexity of the atomistic (molecular) description of the system through the use of a coarse-grained model and concept of beads. In this method, a number of atoms or even group of molecules are combined into particles - beads that interact with each other through soft conservative and pair wise dissipative and random forces. \\
The system evolves according to Newton`s equation of motions with the force on each particle being given by:\\
\begin{equation}
    \boldsymbol{F}_{i} = \sum\nolimits_{i\neq j} (\boldsymbol{F}_{ij}^{C} + \boldsymbol{F}_{ij}^{D} + \boldsymbol{F}_{ij}^{R}),
    	\label{simeq1}
\end{equation}
\noindent acting between given bead $i^{th}$ and its neighboring bead $j^{th}$  in case if the $r_{ij}~<~r_{c}$. Such pairwise interactions allow to preserve the momentum for each pair of beads and to describe the correct hydrodynamic behavior of system and fluid flows.\\
The first term in equation \ref{simeq1} is a conservative repulsive force 

\begin{equation}
    \boldsymbol{F}_{ij}^{C} = a_{ij}\omega(r_{ij})\hat{\boldsymbol{r}}_{ij},
    	\label{simeq2}
\end{equation}
\noindent which is responsible for the bead excluded volume. It represents so-called soft potential characterized by the parameter $a_{ij}$, which can be related to the conventional Flory-Huggins parameters.\cite{Warren1997} The higher the value of $a_{ij}$ the stronger the repulsion between the $i^{th}$  and the $j^{th}$ beads. Such a form of conservative force enables the modeling systems over extended time scales. Second term in  equation \ref{simeq1} is a dissipative force

\begin{equation}
    \boldsymbol{F}_{ij}^{D} = -\gamma\omega(r_{ij})^{2}(\hat{\boldsymbol{r_{ij}}}\cdot \boldsymbol{\nu_{ij}})\hat{\boldsymbol{r}}_{ij},
    	\label{simeq3}
\end{equation}
\noindent representing the effects of viscosity, and the third term is a random force

\begin{equation}
    \boldsymbol{F}_{ij}^{R} = \sigma\omega(r_{ij})(\Delta t)^{1/2}\hat{\boldsymbol{r}}_{ij},
    	\label{simeq4}
\end{equation}
\noindent  associated with thermal fluctuations, $\xi_{ij}$ is a standard normal variable with zero mean. All forces vanish beyond the cutoff radius, $r_{c}$. $\omega(r_{ij})=1 - {r}_{ij}/{r_c}$ are r-dependent weight functions which turns off the drag and random forces at a suitable inter-particle separation $r_{ij} > r_{c}$. $\hat{\boldsymbol{r}}_{ij} = \boldsymbol{r}_{ij}/{r}_{ij}$, $\boldsymbol{\nu}_{ij} = \boldsymbol{\nu}_{i} - \boldsymbol{\nu}_{j}$ is the relative velocity of the beads i and j, and $\Delta t$ is a simulation timestep. Combination of the dissipative and the random forces, serve as heat sink and source and is specified by a friction coefficient $\gamma$ and the noise amplitude $\sigma$. They act as the thermostat in the system. The fluctuation–dissipation relation requires that\cite{Espanol1995}

\begin{equation}
  \sigma^{2} = 2\gamma k_{B}T,
    	\label{simeq5}
\end{equation}

\noindent where $k_{B}$ the Boltzmann constant, $T$ is the temperature.\\
When modeling polymer and solid nanoparticle, the integrity of the chain or the nanoparticle is ensured by an additional spring force between neighboring beads given by

\begin{equation}
   F_{ij}^{bonds} = -k_{bond~ij}(r_{ij} - r_{eq})\hat{\boldsymbol{r}}_{ij},
    	\label{simeq6_new}
\end{equation}

\noindent where $k_{bond~ij}$ is the bond stiffness, and $r_{eq}$ is the equilibrium bond length.\\
To prevent the solid nanoparticles from the deformation at the liquid/liquid interface upon compression, we introduce angle potential which is given by

\begin{equation}
  U_{ijk}^{angles} = k_{bend~ijk}(1 + cos(\Theta)),
    	\label{simeq7}
\end{equation}
\noindent where $k_{bend}$ is the bending stiffness, and $\Theta$ is the angle between two pairs of connected beads sharing a common bead.\\
All quantities in the system are measured in units of the mass of the bead, $m$, thermal energy, $T$, and cutoff radius of the interaction potential, $r_{c}$. Without loss of generality, we set $m = k_{B}T = r_{c} = 1$. Consequentially, our unit of time $\tau = r_{c}(m/k_{B}T)^{1/2}$ is equal to 1. The noise amplitude $\sigma$ in the simulations was 3.0. Density $\rho$ = 3.0. The equations of motion are integrated in time with a modified velocity-Verlet algorithm with a time step $\Delta t$ = 0.01.\\
There are four different types of beads in the system: water (W), oil (O), the beads forming NIPAm microgel shell (M) and solid core particle (P). The molecular structures and coarse-grained approach for the simulated NIPAm based core-shell microgels at the oil/water interface are summarized in the section “Parameters” in SI. Based on the chemical structure and using water-mapping scheme we estimated the characteristic size and molecular weight of the single bead, $r_{c}$= 0.93~nm and $m$ = 162~Da. Thus, 1 decan molecule, 18 water molecule and 1.5 NIPAm monomers is represented by single bead of type O, W and M correspondingly. The properties of all types of beads are summarized in Table~S8 in SI. The interaction parameters between beads of the same species is set to be $a_{ij}$ = 25 while the cross-species parameters reflect the chemical properties of the components. The repulsion parameters for decane–water, microgel–decane, microgel–core, decane-core and water-core pairs were defined using Hansen solubility parameters\cite{Hansen2007}, while the parameters for pNIPAm-water interaction, we obtained based on the approach proposed by Balazs et al.\cite{Yong2013} for simulating thermoresponsive gels. The detailed information about the estimation of the interaction parameters is presented in the SI.\\
\textbf{Synthesis} Three core-shell ellipsoidal microgels CS-L, CS-M and CS-S with different shell characteristics have been prepared. The core-shell microgels were designed as follows. We constructed a unit cell of the diamond crystal lattice where the vertexes correspond to tetrafunctional crosslinkers. Then we assemble two cubic supercells S50 and S25 consisted of 50$\times$50$\times$50 and 25$\times$25$\times$25 unit cells respectively. Supercell S50 is considered as a template for the solid nanoparticle, while S25 – as a template for the polymeric shell.\\ 
The dimensions of the hematite-silica core in the experiment were 330~$\pm$~12~nm $\times$ 75~$\pm$~8~nm. We created the ellipsoidal nanoparticle having a similar ratio between the long and short semi-axis 4.4$\times$1. The ellipsoidal nanoparticle was constructed by inscribing the ellipsoidal-shaped frame into S50 supercell and cropping all the beads, which are outside of the ellipsoid. Beads forming the nanoparticle are denoted by P. Similar to the experiment we used the same size nanoparticle for all types of microgels.\\
To construct the polymeric shell around the solid nanoparticle we used a scaled S25 supercell. All the bonds between the tetrafunctional crosslinker were replaced by the subchains of different lengths. The distribution of subchain lengths is described by a Gaussian distribution with a mean value equal to 11 and standard deviation equal to 2. Two ellipsoidal frames of different radii were inscribed into the modified supercell. The small frame is necessary for the formation of the void in the polymeric network of the same sizes as the solid nanoparticle ones. The sizes of the large frame control the thickness and the shape of the polymeric shell. All beads, which were located inside the small and outside of the large ellipsoidal frames, were cropped. The rest of the beads forming microgel shell were denoted as M. Such procedure allowed us to create a hollow microgel with uniform distribution of cross-linkers and the average cross-link density equal to 4.5\%. Then the solid nanoparticle was inserted into the void of the microgel with further grafting of the dangling chains of the polymeric shell to the nanoparticle surface. The dangling chains of the polymeric shell were physically attached to the nanoparticle surface. Going through all of the free ends of the dangling chains we monitor whether the distance between it and closest bead at the surface of the solid core satisfies the condition: $|r_{end}-r_{surf}|\leq r_{c}$. If so, a bond between the free end of the dangling chains and the bead of the surface of the solid core was formed. Only one bond could be formed between the free end and the solid core. The thickness of the shell was controlled by the radii of the large frame. We tune the radii in such a way as to reproduce the experimental ratio between the sizes of the core and the shell of the CS-356, CS-254\cite{Nickel2019}, and CS-165 microgel in equilibrium in the swollen state (Table~S2, S3). Microgels CS-L, CS-M and CS-S with thick, medium and small shell were created. The number of beads, bonds, and angles are presented in Table S4 - S6. Independently of the shell thickness, 144 extra bonds were created after attaching the shell to the solid core.\\
\textbf{Microgels in bulk}\\
The simulations of single microgels in water solution were performed in the NVT ensemble in cubic boxes of a constant volume V = $L_{x}$ $\times$ $L_{y}$ $\times$ $L_{z}$ = 100 $\times$ 100 $\times$ 100 $r_{c}$ with periodic boundary conditions in all directions. The necessary amount of water beads, $W$, were added to the simulation box. All the microgels were equilibrated. The equilibration of the system lasted for 2·10$^{6}$ steps.\\
\textbf{Microgels at the oil/water interface}\\
By analogy with experiments, we prepared three types of monolayers, consisted of either CS-L, CS-M or CS-S microgels. The microgel monolayers were formed by 16 microgels placed at the oil-water interface (Figure~S9). The initial size of the simulation was $L_{x}$ $\times$ $L_{y}$ $\times$ $L_{z}$ = 240 $\times$ 210 $\times$ 70 $r_{c}$. This size was used to reproduce the procedure of the Langmuir Blodgett trough within the experimental part. Different simulation boxes were tested to investigate the effect of the simulation box geometry (see supporting information). The ratio between the numbers of water and oil beads was 2:1 and the total number of beads in the simulation box was more than 10$^{7}$. 16 swollen microgels have been located near the water/oil interface. Each of the microgels was tilted with respect to the OZ axis to the angle of 50°.  Moreover, all the microgels were rotated to the random angle around OZ to eliminate the possibility of introducing the order into the monolayer by the initial conditions.\\
After the first equilibration for 2.5 $\times$ 10$^{6}$ steps, the systems were compressed along the x-axis while preserving its volume ($L_{y}$ was fixed and $L_{z}$ increased). We perform $N$ = 9 cycles of compression. On each cycle, we gradually decrease the length of the simulation box by 20$r_{c}$  within the 0.5$\times$10$^{6}$ timesteps. After the compression, the system was equilibrated during the 2.5$\times$10$^{6}$ timesteps and the next stage of compression began. We repeat such procedure until $L_{x}$ = 60$r_{c}$.
We applied several criteria to judge the equilibrium in a system, including the equilibrium of energy, the equilibrium of gel sizes, and the equilibrium of order parameter. We waited until values of all of these characteristics reached a plateau, i.e. when they become invariable or have small fluctuation around the average value. The example of the variation trends of the order parameter with simulation time for the CS-M microgels are shown in Figure~S13.

\section{Results and Discussion}
\begin{figure}[ht]
\centering
\includegraphics[width=0.75\linewidth, trim={0cm 0cm 0cm 0cm},clip]{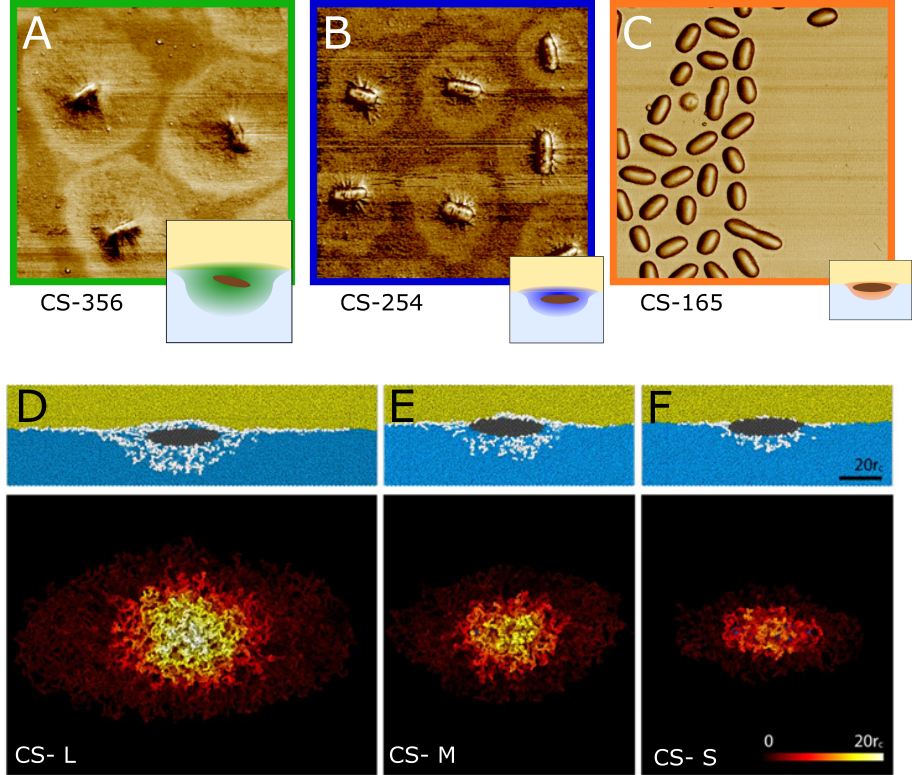}
	\caption{Phase images (AFM) of the hybrid microgels CS-356 (A), CS-254 (B) and CS-165 (C) and their respective sketches to illustrate the shell characteristics. All images are 3~µm~$\times$~3µm. Snapshots of the single microgels (side and top views) with different thicknesses of the shell CS-L (D), CS-M (E) and CS-S (F) at the oil-water interface: the cross-sections of the thickness of $d = 2r_{c}$ near the center of mass (along the long axis of the core) and surface structure (height maps) of the adsorbed microgels (view from the water side).}
	\label{AFM_3mum}
\end{figure}

\textbf{Characterization of microgels}\\
\textbf{Experiment}\\
In this study, three hybrid microgels with different shell characteristics are used to investigate the influence of the shell composition on the behavior at the decane-water interface. All three microgels are synthesized on the same batch of hematite-silica cores and differ only in the microgel shell added via seed and feed precipitation polymerization. The dimensions of the hematite-silica core of approximately 75~$\times$~330~nm were obtained by TEM.\cite{Nickel2019} The shell sizes differ with respect to the shell-thickness and additionally their softness. The different shell thickness in the collapsed state is a result of the different monomer-to-core mass-ratio used for the respective syntheses (see SI). The softness can be influenced by the co-monomers and cross-linker.\\
In general, information about microgel dimensions and softness in bulk solution is obtained via the hydrodynamic radii obtain with dynamic light scattering.\cite{Dubbert2014,Brugnoni2018} For anisotropic microgels, an apparent hydrodynamic radius can be calculated. This radius gives an approximation about the shell sizes and their swelling ratios. Thus, it is possible to obtain information about the size and structure of the microgels in the solution. Relying on this information, we can forecast size, structure and anisotropy of the microgels at the interface: Microgels with a larger shell and higher swelling degree will deform stronger and occupy a larger area at the interface. As a result, such microgels will be less anisotropic at the interface as compared to their anisotropy in bulk solution.\\
The first microgel, CS-356, is the largest microgel as it has the largest shell. DLS gives a hydrodynamic radius of 356~$\pm$~10~nm in the swollen state (10~°C) and the hybrid microgel deswells to 162~$\pm$~2~nm in the collapsed state at 50~°C (see SI). The second microgel, CS-254, shows a decrease from 254~$\pm$~4~nm to 114~$\pm$~1~nm in the hydrodynamic radius (see SI). Hence, CS-254 has a similar swelling degree as CS-356, yet, it is much smaller. The last microgel, CS-165, is stiffer compared to the other two microgels. From DLS a swelling degree of 1.4 is determined, as the microgels show a hydrodynamic radius of 165~$\pm$~3~nm in the swollen and 122~$\pm$~1~nm in the collapsed state. This swelling degree is much smaller compared to the swelling degree of 2.2 for the prior two microgels. Similar effects are observed in two dimensions at the interface.This large effect on the swelling degree is not expected as the amount of available cross-linker was only increased by approximately 1~mol\%. Nevertheless, the shell of CS-165 is stiffer compared to the other two as can be seen by the smaller swelling ratio obtained in the DLS and the reduced spreading at the interface. Yet, there is no reason that the small amount of BAC changes the surface properties. The introduction of BAC to the microgel network does not introduce additional charges. Hence, the main difference between all three core-shell microgels is their shell thickness and an additional difference is the stiffer network of CS-165 as compared to the other two.  \\
The hydrodynamic radii in the collapsed state are in agreement with the shell proportion obtained with thermogravimetric analysis (see SI). The shell of CS-356 is 77.3~\% of the total weight of the hybrid core-shell microgel and has the largest collapsed radius. CS-254 and CS-165 have a similar shell weight of 60.9~\% and 65.0~\% of the total weight of the hybrid core-shell microgel respectively and additionally similar collapsed radii. The difference within the swelling degree is a result of the different co-monomer \textit{N},\textit{N}’-bis(acryloyl)cystamine (BAC) used for the synthesis of CS-165 compared to the dye methacryloxyethyl thiocarbamoyl rhodamine B (MRB) used for the other two syntheses.\\
In Figure~\ref{AFM_3mum}~A-C, AFM phase images of 3~µm~$\times$~3~µm in size demonstrate the three investigated microgels deposited onto a solid substrate. Additionally, sketches of the microgels are shown to visualize their appearance at the decane-water interface. CS-356 consists of the anisotropic hard core surrounded by a largely spread microgel shell (Figure \ref{AFM_3mum}~A). The shell contains a higher cross-linked area located directly at the core and a less cross-linked outer part, resulting in a thin microgel film in 2D. The spreading of the microgel shell is larger at the sides of the core when compared to the tips of the core, resulting in an almost spherical shape. Figure \ref{AFM_3mum}~B shows the smaller shell of CS-254 in 2D. Similar to CS-356, the shell of CS-254 is spreading in an almost spherical shape. CS-165 appears anisotropic in 2D (Figure~\ref{AFM_3mum}~C). The microgel shell is stiffer, hence, no spreading of the shell is observed. As a result, CS-165 shows an apparent aspect ratio of 1.8 in 2D. In few cases, one core-shell microgel has two elliptical cores incorporated as the existence of double core microgels can only be reduced with sonification but not prevented completely. These microgels were neglected for the calculation of the aspect ratio.

\textbf{Simulations}\\
Computer simulations of single microgels in water solution and at the oil-water interface were performed. Three core-shell ellipsoidal microgels CS-L, CS-M and CS-S with different shell thicknesses have been prepared. All the samples contain identical solid cores represented by the ellipsoidal nanoparticle. The aspect ratio (the ratio of the lengths of the long axis to the shorter one) of the solid nanoparticles was 4.4$\pm$1 which coincide with the dimensions of the hematite-silica core in experiment. The microgel with the thickest shell is denoted CS-L, while the microgels with thin and the thinnest shell are denoted CS-M and CS-S, respectively.\\
At T=20~°C, all microgels are in a swollen state in the bulk water. The concentration profiles show that the microgels contain approximately 80\% solvent (Figure S8). To perform accurate comparison between the experiment and simulations, the relative sizes of the microgel were calculated. The information about the lengths of the major semi-axis $L$ and minor semi-axis $R$ of the microgels in a swollen state in solution is obtained using the LSQR approximation of a surface of the microgel shells and via the calculation of gyration tensor (Table S2 and Table S3). We normalized $L$ and $R$ radii of the microgels to the length of the small principal axis of the solid core, which is independent of the system. In case of the microgels with large and medium shell, the aspect ratio $L_{bulk}/R_{bulk}$ = 1.8$\pm$0.04 and 1.78$\pm$0.03 were found which is in a good correlation with experiment. Thus, one could confidently state that microgels CS-L and CS-M relates to the microgels CS-356 and CS-254 respectively. The microgel with the thinnest shell, CS-S, has a more anisometric shape, compared to CS-L and CS-M. One should notice that the CS-S microgel was assumed to have the same type of beads and have the same cross-link density as the other samples in contrast to the experiment. Hence, the direct comparison between CS-S and CS-165 is not possible as CS-165 has a smaller swelling degree compared to CS-356 and CS-254 although that we have used same proportion of cross-linker BIS in the synthesis. With simulations, it is possible to investigate directly the effect of the shell thickness without changing any other parameter of the shell, while within the experimental work the exchange of the dye methacryloxyethyl thiocarbamoyl rhodamine B (MRB) to BAC does not only add 1~\% more cross-links but may change the polymerization kinetics resulting into a different distribution of cross-links and a stiffer microgel shell.\\
The snapshots of the equilibrium structure of the three investigated microgels adsorbed at the oil/water interface are presented in Figure \ref{AFM_3mum}~D-F. We estimated the large $L_{int_{0}}$ and small  $R_{int_{0}}$ radii of the microgel in plane, which is parallel to the interface, and the radius of the microgel in the normal direction, $h$ (Table \ref{tab1aSimMP}). We observe significant spreading of the microgel shell at the interfaces. Due to the high incompatibility between the liquids, the microgel tends to screen unfavorable contacts between water and oil beads which lead to increasing the lateral sizes of the microgels (Table \ref{tab1aSimMP}). All microgels have a lower size anisotropy at the interface compared to their anisotropy in bulk solution (Table \ref{tab1aSimMP}). \\
Comparing the relative size changes of the polymer shell at the tips, $l = L_{int_{0}}/ L_{bulk}$, and at the sides, $r =  R_{int_{0}}/ R_{bulk}$, of the core, we found that $l_{CS-L}$ > $l_{CS-M}$ > $l_{CS-S}$ and $r_{CS-L} > r_{CS-M} > r_{CS-S}$. Microgels with a larger shell deform stronger and occupy a larger area at the interface. At the same time, oil is a poor solvent for the polymer shell thus the microgel tends to optimize the contact area by reducing anisotropy to make it more circular, which is in agreement with the experimental findings. Comparing the relative size changes between $l$ and $r$, we found that $r > l$ for all the samples. It means that the spreading of the polymer shell is larger at the sides of the core when compared to the tips of the core.\\
The height maps of adsorbed microgels are shown in the bottom row of Figure \ref{AFM_3mum}. The zero level of all of the maps is the position of the oil/water interface. Each map has the same color scales. We observe different degrees of immersion of the microgels into the different liquids.  Weak solubility of the polymer shell in the oil leads to the swelling of the microgel into the aqueous phase. The swollen microgel in the water phase is surrounded by the flat, strongly adsorbed shell. The position of the solid core depends on the thickness of the shell, which is similar to the observations made for spherical core-shell microgels from Vasudevan et al.\cite{Vasudevan2018} However, for our anisotropically shaped microgels the vector passing through the major axis of the solid particle was almost parallel to the plane of the oil/water interface for all samples. It is worth bearing in mind that experimentally the shape and size of microgels are determined at the solid/air interface after the deposition of the microgels onto a solid substrate.  The air can be treated as a bad “solvent” for the microgels. In this case, polymeric shells collapse, adhere to the surface of the particle and spread more strongly on the hard surface.

\begin{table}[H]
     \centering
    \caption{Gyration radii of the microgel adsorbed at oil/water interface at T=20~°C. $L_{int_{0}}$ and $R_{int_{0}}$ the large and small radii of the microgel in the plane, which is parallel to the interface, $h$ is the size of the microgel in the normal direction.}
     \begin{tabular}{c|ccc|c|c|c}
       \hline
      Gel &	$L_{int_{0}}$	& $R_{int_{0}}$	&$h$	&$L_{int_{0}}/ R_{int_{0}}$&	$r = R_{int_{0}}/ R_{bulk}$&	$l = L_{int_{0}}/ L_{bulk}$ \\
\hline
CS-L &	23.24$\pm$0.15 &	13.64$\pm$0.14&	4.37$\pm$0.09 &	1.70$\pm$0.01 &	1.81$\pm$0.03 &	1.58$\pm$0.02\\
CS-M &	15.21$\pm$0.15 &	9.63$\pm$0.11 &	3.61$\pm$0.08 &	1.56$\pm$0.02 &	1.62$\pm$0.03 &	1.38$\pm$0.02\\
CS-S &	11.50$\pm$0.12 &	5.96$\pm$0.07 &	2.68$\pm$0.06 &	1.92$\pm$0.03 &	1.51$\pm$0.03 &	1.22$\pm$0.02\\
       \hline
     \end{tabular}
     \label{tab1aSimMP}
     \end{table}

\begin{figure}[ht]
\centering
    \includegraphics[width=1\linewidth, trim={0cm 0cm 0cm 0cm},clip]{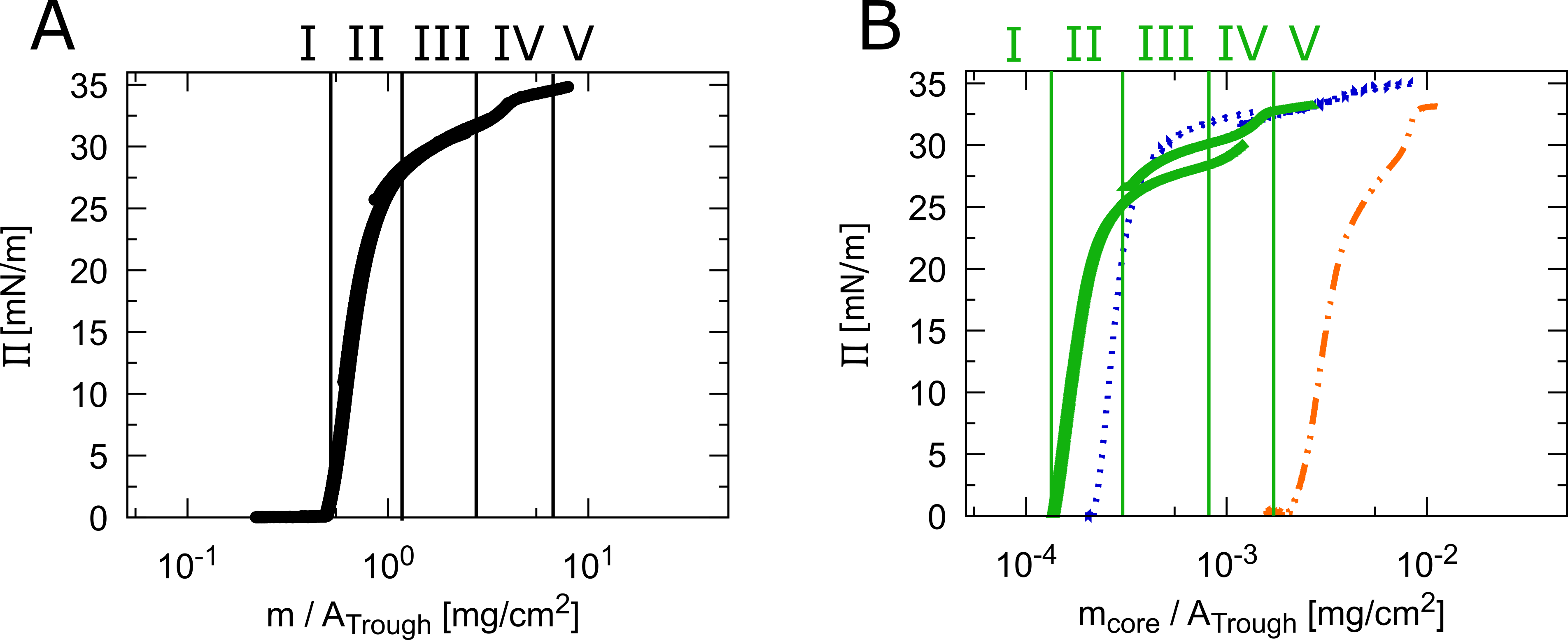}	
	\caption{A: Typical compression isotherm of a regular microgel shown with the corresponding five regions. The image is adapted from Bochenek et al.\cite{Bochenek2019} B: The compression isotherms in dependence of the core mass divided by the available trough area for CS-356 (green, solid), CS-254 (blue, dotted) and CS-165 (orange, dashed) show their different reaction by an increase in surface pressure with differently shaped isotherms. The five different regions are highlighted for CS-356.}
	\label{IsothermImage}
\end{figure}

\textbf{Microgel monolayer}\\
\textbf{Experiment}\\
Compression isotherms are important to obtain first information about their different compression states at the interface. A comparison of the compression isotherms is shown in Figure \ref{IsothermImage}~B for CS-356 (green, solid line), CS-254 (blue, dotted line) and CS-165 (orange, dashed line). The compression isotherms are normalized for the mass of the cores added to the interface. The weight of the core within each of the microgels is obtained by thermogravimetric analysis (for further information see SI). This scaling is used as each microgel was synthesized on the identical batch of cores. By normalizing with the mass of the core, we obtain the same number of particles and are able to show the compression isotherms with respect to the same microgel number applied to the interface. By using this approach, we neglect the amount of microgels not adsorbed at the interface. Nevertheless, the approach of normalizing the compression isotherms with the number of visible particles in the AFM images\cite{Bochenek2019} is not suitable in our study due to the clustering of CS-165 which will be discussed later.\\
The large shift of the compression isotherm of CS-165 compared to the two other microgels can have two reasons: First, the microgels cover a smaller effective area fraction at the interface as the microgel shell does not spread as much as compared to the two microgels with the higher swelling degree. This could cause weaker adsorption of the microgels to the decane-water interface. As a result, some microgels may not attach to the interface and sink into the water phase during the preparation of the monolayer. Second, the microgels occupy a smaller area at the interface and, consequently, a higher number of microgels compared to the larger microgels are needed to induce an increase in the surface pressure. Nevertheless, the shapes of the three compression isotherms are not influenced by shift on the x-axis. In this study, we focus on the different shapes of the compression isotherms as this gives more information about the structures and sequence of the different monolayers.

\begin{figure}[H]
\centering
    \includegraphics[width=1.0\linewidth, trim={0cm 0cm 0cm 0cm},clip]{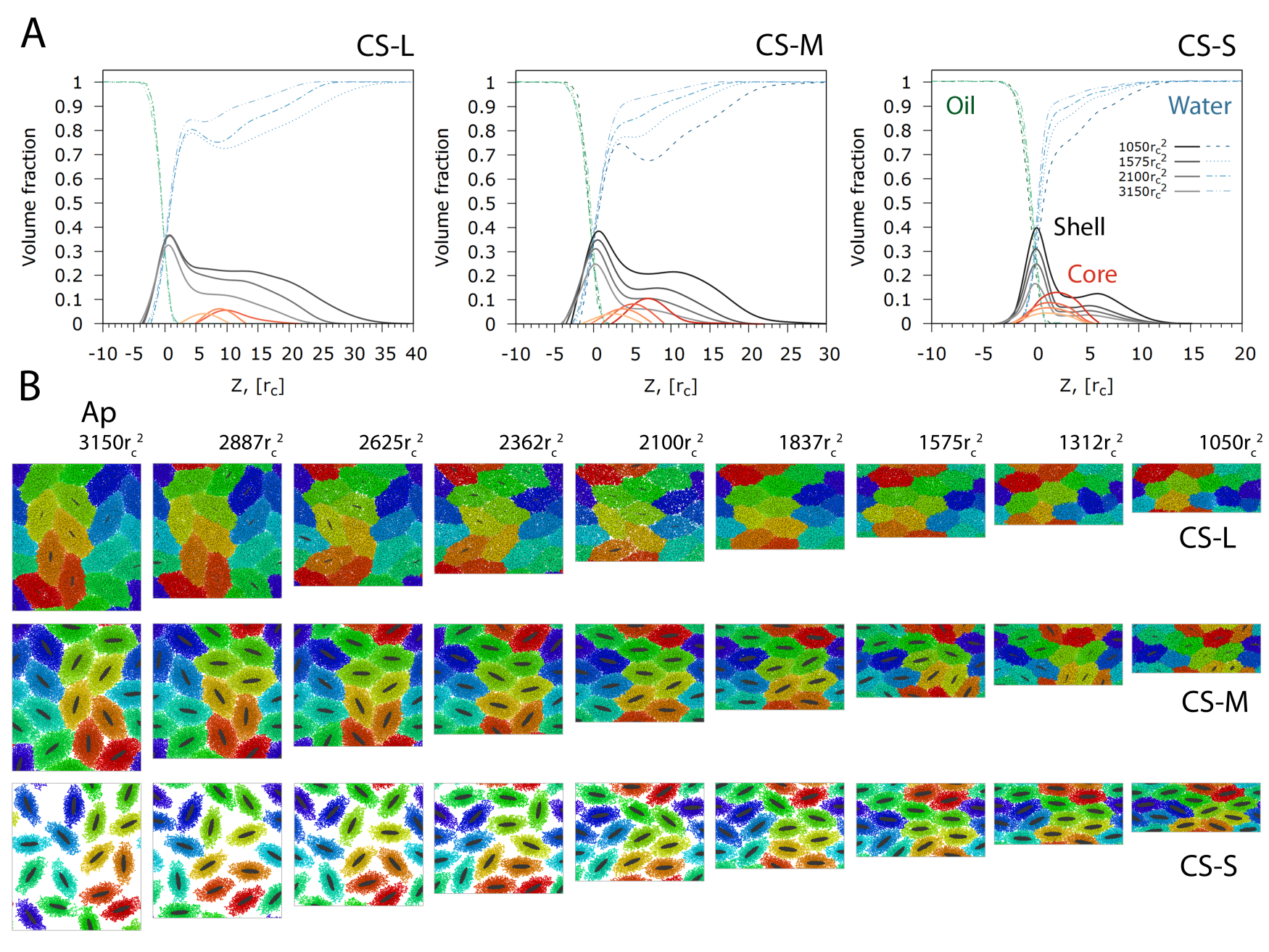}	
	\caption{A: Concentration profiles along the normal (z-axis) to the liquid interface within adsorbed microgels. Grey, orange, blue and green curves  correspond to the distribution of beads in the shell, solid core, water and oil, respectively. The microgels have different shell thicknesses: CS-L, CS-M and CS-S. Dark to light (shell and core) gradient color scheme and dashed to dotted  lines (water and oil) correspond to the cases of high, $Ap$ = 1050 $r_{c}^{2}$, medium, 1575 $r_{c}^{2}$, low, 2100 $r_{c}^{2}$, and weak, 3150 $r_{c}^{2}$, compressions, respectively. Z = 0 corresponds to the position of the water/oil interface. B: The snapshots of slices of microgel monolayers for different compression degrees (surface coverage). For clarity, the polymer shell of different microgels is marked by rainbow colors, while solid cores are in gray. The slice of thickness 3$r_{c}$  was made at the plane of the water/oil interface.}
	\label{Simulation1}
\end{figure}

Compression isotherms of spherical microgels reveal five different regions as a result of their core-corona structure which are highlighted in Figure~\ref{IsothermImage}~A.\cite{Bochenek2019} \cite{Geisel2015} Bochenek et al.\cite{Bochenek2019}  determined the five different regions within the compression isotherms based on different phases observed on the respective AFM images. Nevertheless, for core-shell anisotropic microgels different phases within the AFM images are not as clear, hence, the designation of the regions was done based on the shape of the compression isotherms and its observed deflections (Figure~\ref{IsothermImage}~B and S4) The microgel with the largest shell shows five regions in the compression isotherm (green) as well. These are less distinct compared to the spherical one. As a guide to the eye, the rough regions for CS-356 are highlighted in Figure~\ref{IsothermImage}~B. At the contact point, the surface pressure increases drastically marking the transition from region I where the microgels are not in contact to region II where the microgels are in contact. In this region, only the microgel coronae are in contact. A pseudo-plateau is observed for the microgels when the available surface is decreased even further (region~III). In literature\cite{Geisel2014b, Rey2016, Bochenek2019}, this pseudo-plateau is explained by the isostructural phase transition between two hexagonal phases for spherical microgels: Microgels where the fuzzy corona of the microgels are in contact and compressed microgels which form a hexagonal phase with smaller distances between the microgels. The microgels are compressed further and more and more microgels adjust to the close hexagonal phase. At the point where all microgels are in core-core contact, a further decrease in the surface area results into a steep increase in surface pressure (region~IV). Compressing the surface area even further leads to the collapse of the monolayer in region V (Figure~\ref{IsothermImage}~A). The observation that the microgel with the largest shell shows a similar isotherm as was reported for spherical microgels\cite{Bochenek2019} and spherical core-shell microgels\cite{Geisel2015} indicates that the spreading of the shell at the interface leads to a circular, isotropic shape at the interface. Consequently, the core-shell microgel, which is anisotropic in dispersion, behaves as an isotropic microgel at the interfaces, although the core is anisotropic. \\
The compression isotherms of the other two microgels differ more from the behavior of microgels with spherical, isotropic shape. For the second microgel (CS-254), the compression isotherm has a less pronounced pseudo-plateau as region III as compared to the previously mentioned microgel and a reduced slope for the rise in surface pressure in region IV. As a result, there is a change of the slope within the compression isotherm marking the transition from region III to region IV. This transition is less pronounced as compared to the transition of region III to IV for CS-356. Region V is again similar to CS-356 and has a slight increase and no plateau. The less pronounced pseudo-plateau of CS-254 can be expected because the thinner shell makes the microgels less compressible compared to CS-356. Figure~S4 in the Supporting Information highlights the five different regions for CS-254. \\
For the last microgel (CS-165), an anisotropic shape in 2-dimensions is visible in the AFM image. Accordingly, it is comprehensive that the compression isotherms differ the most from that of a spherical microgel: Region III shows a slight decrease in the slope of the compression isotherm. Nevertheless, the surface pressure rises within region III of the microgel CS-165. The increase of the surface pressure in region IV of the compression isotherm for CS-165 is much steeper as compared to the other two. Nevertheless, region V displays a plateau similar to spherical microgels, indicating buckling of the monolayer. Figure~S4 in the Supporting Information illustrates the transitions between those five different regions for CS-165.\\
\textbf{Simulations}\\
The advantage of using DPD simulations to study the compression behavior of the hybrid core-shell microgels monolayers at the water/oil interface is that the underlying structural details can be explored. Figure S10 provides an overview of the simulated 3D height images of monolayers at different compression degrees for the three different hybrid core-shell microgels. The maximum height corresponds to the protrusion of the microgels into the water phase. To characterize the compression degree, we use the parameter $Ap$, which is the ratio of the total interfacial area to the number of the microgels at the interface. At large $Ap$, before overlapping at region I, microgels are spaciously positioned on the surface and barely interact with each other. The microgels are in an undisturbed state and behave similarly to the case of single gels discussed above. The measurement of the normal concentration profile of the monolayer demonstrates that the maximum of the polymer concentration is near the interface (z~$\sim$~0), which occurs due to the effective shielding of unfavorable contacts between the liquids (Figure \ref{Simulation1}~A). The same behavior was also observed recently for spherical microgels.\cite{Bushuev2020} The solid particles are located directly near the interface, in the part of the microgels that swells into the aqueous phase. The distribution of directions of the main axis of solid cores is random.

\begin{figure}[ht]
\centering
    \includegraphics[width=0.89\linewidth, trim={0cm 0cm 0cm 0cm},clip]{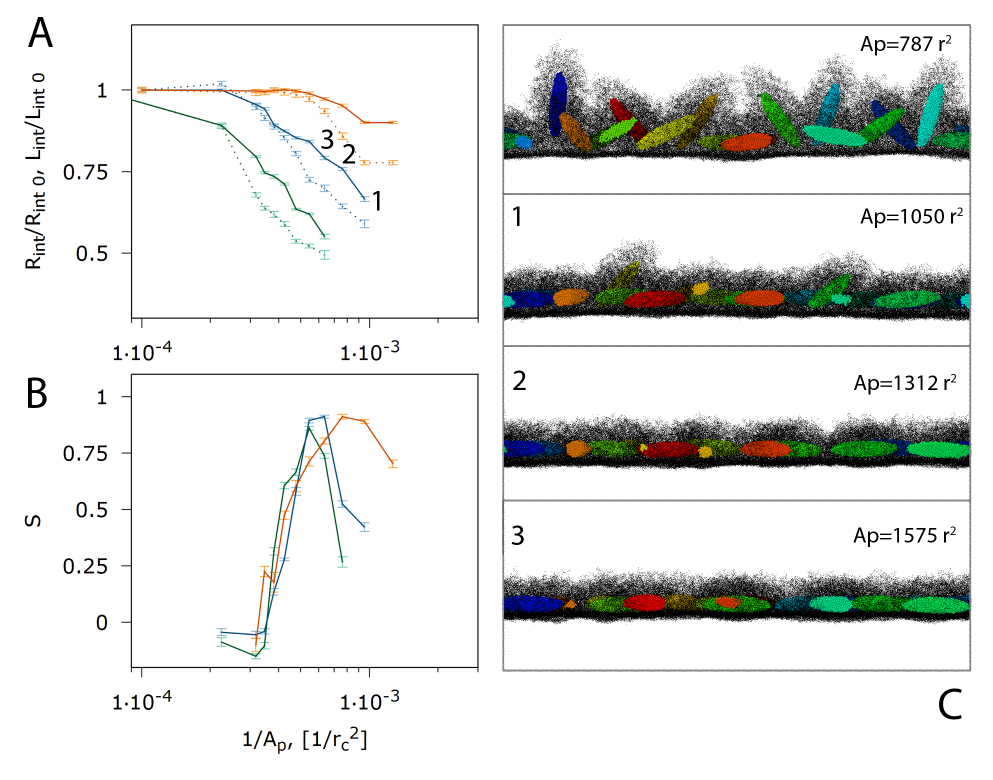}	
	\caption{A: Normalized values of components of the mean gyration radii of single microgels in the monolayer as a function of inverse area per particle of the microgel-covered oil-water interface. Solid and dashed lines correspond to the mean values of the long and short radii of the spreading area of the microgel. B: Evolution of the 2D order parameter for the solid cores as a function of inverse area per particle of the microgel-covered oil-water interface. The curves of the different colors correspond to the microgels of different shell thicknesses: CS-L (green), CS-M (blue) and CS-S (orange). C: Side view snapshots of CS-S microgels at 1575~$r^{2}$, 1312~$r^{2}$, 1050~$r^{2}$ and 787~$r^{2}$.}
	\label{Simulation2}
\end{figure}

To understand how the compression of the monolayer influences the deformation of the whole particles, we calculated the main components of the mean gyration radius of microgels. Moreover, to estimate the spreading of the microgel at the interface we also perform a calculation of 2D gyration radius of the thin layer of thickness 3$r_{c}$ at the plane of the water/oil interface, -1$r_{c}$<Z$\leq$2$r_{c}$. This is a layer where the polymer concentration reaches the peak on the concentration profile. Figure \ref{Simulation1}~B provides an overview of the corresponding region for CS-L, CS-M and CS-S at different compression states.\\
Normalized values of components of the mean 2D gyration radii of single microgels in the monolayer as a function of inverse area per particle of the microgel-covered oil-water interface are plotted in Figure \ref{Simulation2}~A.  The solid and dashed lines depict the relative change in long, $L_{int}/L_{int_0}$, and short, $R_{int}/R_{int_0}$, radii of the adsorbed microgels, where $L_{int_0}$ and $R_{int_0}$ are the large and small radii of the undeformed microgel. One could observe the general decrease of the overall size of each microgel upon monolayer compression. However, the relative size changes of the polymer shell at the tips and the sides of the solid core are different. Under compression, microgels deform more strongly at the sides of the core than the tips. The only exception is the behavior of CS-L and CS-M microgel at the region of high area per particle $Ap$. After the crossing point where the microgels start to interact with each other (region II) microgels deform more strongly at the tips of the core than at the sides. The results of relative anisotropy of the microgels $L_{int}$/$R_{int}$ as a function of area per particle are shown in Figure S11. We observe that compression of the monolayer at first decreases the anisotropy of the microgels at the interface, while as the monolayer is compressed, the anisotropy of the microgels significantly increases. The increase of the compression increases the maximum height (protrusion into the water phase) of the monolayer, increases the polymer concentration, and normal size of the microgel in water (Figure \ref{Simulation1}~A). Z position of solid particles shifts more into the aqueous phase.  What is more important, there is a dependence of directions of the main axis of solid particles on the compression degree, which will be discussed further below.\\
In case of the microgels having the thick shell, CS-M and CS-L we also observe the tilting of the solid particles with respect to the interface. As the monolayer is compressed (region IV-V), some of the solid particles have a tendency for the inclitaion (Figure S10), increasing the width as well as the inhomogeneity of the concentration profiles and thus the thickness of the monolayer. At the highest compression levels $Ap$ < 1050 $r_{c}^{2}$, the concentration profiles of CS-L system are very broad, caused by the monolayers buckling and collapsing. The desorption of the CS-L microgels is observed.\\
Thus, independent of the shell thickness, a compression of the monolayer is accompanied (i) by gradual increase of the polymer concentration, (ii) by thickening of the monolayer into the aqueous phase, (iii) by an increase of the anisotropy of the microgels in the lateral direction, (iv) by a rotation of the solid cores around the normal perpendicular to the interface, and (v) by shifting the position of the solid nanoparticles away from the interfaces deeper to the aqueous phase. 

Now, let us demonstrate how the ordering of the microgels in the monolayer changes upon its compression in simulations. To provide a quantitative measure of the degree of orientational order in the system, we used the scalar order parameter $S$ defined in SI. It takes the value 0 in the disordered state and the maximum value 1 in the complete ordered state. The description of the orientation order in the system is based on the analysis of the directions of the long axis of the solid cores in the microgels. In 2D, we analyzed the projection of the tilted core long axis onto the xy-plane of the water/oil interface. The result is given in Figure \ref{Simulation2}~B showing the evolution of the order parameter as a function of the inverse area per particle $Ap$. Initially, at the region I and at the beginning of the region II, there is no ordering of solid cores, $S$=0. As demonstrated in Figure S12, all the system show a homogeneous disordered isotropic state where all directions of the long axis of the solid cores are equally probable.\\
Starting from the end of the region II, the disorder to order transition occurs with a jump in the order parameter. In case of CS-L and CS-M microgels we observe the peak of the order parameter in region III and then order parameter gradually decreases in region IV and V. In case of the microgels of the smallest shell, we observe the shift of the peak to low $Ap$ values (region IV and V).\\
The microgels show a transition from a disordered state at high available areas, $Ap$, to a more ordered state at less $Ap$, reverting to a less ordered state again at even lower $Ap$. This is related to the peculiarities of changes in the anisometry of the microgel shells as a reaction of the compression and the pursuit of the system to minimize the oil/water contact area (Figure \ref{Simulation2}~A) as well as the mediated interaction of the solid cores.

At high available areas (region I), there are no prerequisites for ordering of the microgels. There is enough free available space for the microgels at the interfaces (Figure \ref{Simulation1} CS-S). At low compression degree, the decrease in $Ap$ leads to the redistribution of the microgels to cover a maximum area at the air/water interface. When the microgels get in contact with each other, their anisotropy decreases, however it does not influence the $S$ value of the core. Looking to the top (Figure~\ref{Simulation1} B) and the side view (Figure~\ref{Simulation2} C) of the systems one can see that all cores are disposed in the plane of the air/water interface and have random orientation. Further compression leads to a rotation of the solid particles in the plane of the air/water interface caused by the interaction, interpenetration, and deformation of the polymeric shells. It is accompanied by a significant increase of the microgels anisotropy, and by the ordering of the solid core. The director direction, which specifies the average local orientation, is almost perpendicular to the direction of compression ($L_{x}$ axis). This increase in the orientational ordering is independent from the used simulation box (see supporting information). In case of equal $L_{x}$ and $L_{y}$ values, an increase in orientational ordering is accompanied by a random orientation of the director with respect to the axes.
At the highest compression, the polymeric shell is significantly compressed and the presence of the solid core comes to the forefront breaking the order. The limitation of the compression of the shell promotes the inclination of the solid particles (Figure~\ref{Simulation2} C and Figure S10) accompanied by the decrease of the orientational order parameter. Further compression leads to the destruction of the monolayer.\\

\begin{figure}[ht]
\centering
\includegraphics[width=0.75\linewidth, trim={0cm 0cm 0cm 0cm},clip]{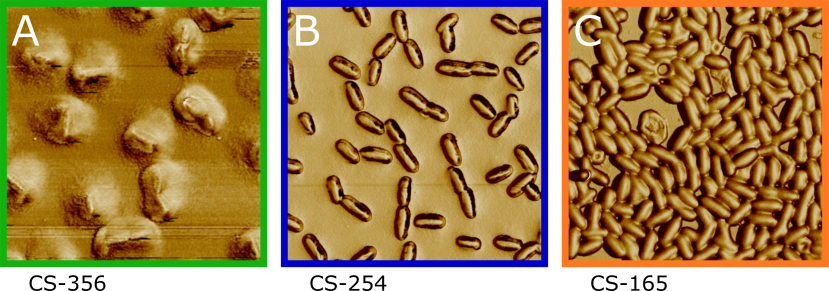}
	\caption{Phase images (AFM) of the three different compressed hybrid microgels. The outer edges of the shells of the different microgels are indistinguishable. The further analysis is based on the visible ellipsoidal cores. All images are 3~µm~$\times$~3µm.}
	\label{AFM_3mumD-F}
\end{figure}

\textbf{Analysis of the experimentally obtained monolayer}\\
Analysis of the AFM images allows interpreting the different shapes of the compression isotherms and to link the changes in the isotherms to the microscopic organization of the microgels in the monolayer. Hence, monolayers were transferred onto silica substrates simultaneous to the compression. These samples were dried and further investigated with AFM. Figure~\ref{AFM_3mumD-F} shows AFM phase images of the three different microgels at higher surface pressure. While the shells are no longer distinguishable, the elliptical cores are clearly visible. The cores are easily accessible in the respective AFM height images and therefore we base the further investigation of the ordering behavior on the elliptical cores and their position, distance and angle to each other.

\begin{figure}[ht]
\centering
    \includegraphics[width=1\linewidth, trim={0cm 0cm 0cm 0cm},clip]{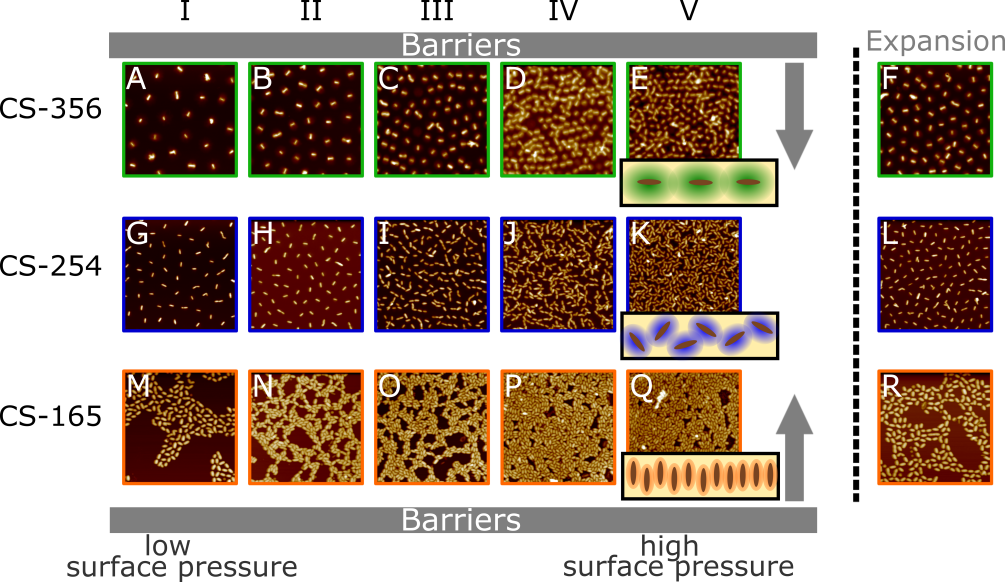}	
	\caption{AFM height images of the three different microgels at each compression region show the variety of monolayer structures. A representative AFM image is shown for each region in the compression isotherm and for each microgel. For region V, sketches show the different ordering of the different microgels: perpendicular (E) or parallel (Q) to the direction of the compression or not ordered at all (K). The direction of the compression is demonstrated with the barriers. Additionally, the images F,L and R reveal AFM height images of the monolayers after expansion. (7.5~$\times$~7.5~µm)}
	\label{MainImage}
\end{figure}

Figure \ref{MainImage} provides an overview of the AFM height images at different compression states/ surface pressures for the three different hybrid core-shell microgels. Furthermore, the direction of the compression with the barriers is indicated with arrows. For each microgel, one image representing the monolayers occurring in each of the five compression states is shown. Subsequent to the compression of the monolayer with the barriers, the barriers were opened leading to an expansion of the available surface area. Images of these monolayers after expansion are additionally presented to investigate the reversibility of the microgel structures. These images will also be used later to discuss the strength of the attractive capillary interactions between the microgels.

\begin{figure}[ht]
\vspace{0.5cm}
\centering
\begin{minipage}{0.5\textwidth}
\includegraphics[width=0.8\linewidth, trim={0cm 0cm 0cm 0cm},clip]{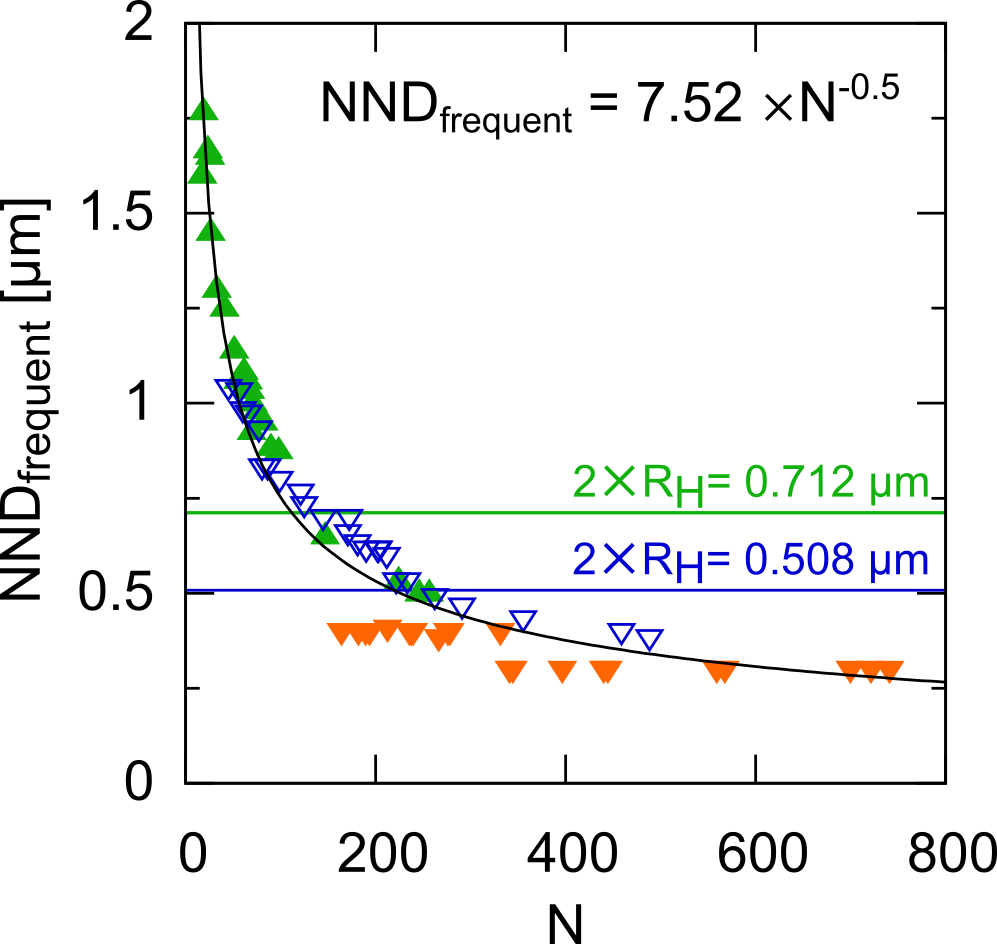}
\end{minipage}\begin{minipage}{0.5\textwidth}
\includegraphics[width=0.8\linewidth, trim={0cm 0cm 0cm 0cm},clip]{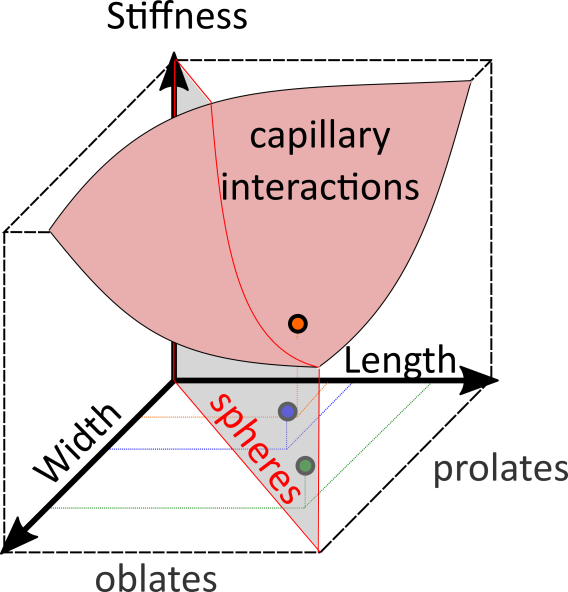}
\end{minipage}
	\caption{Left: Most frequent NND depending on the number of cores is shown for CS-356 (green full pointing up triangles), CS-254 (blue empty pointing down triangles) and CS-165 (orange full pointing down triangles). The black line demonstrates the isotropic connection between NND$_{frequent}$ and N and the blue and green lines identify the sizes of CS-356 and CS-254 in bulk. Right: Sketch of the dependence of the width and length of the particles and their stiffness to the appearance of capillary interactions. The circles in green, blue and orange show the respective position of the three core-shell microgels within this connection.}
	\label{NND_plot}
\end{figure}

The images of CS-356 are in the upper row, ranging from the lowest surface pressure (A) to the highest surface pressure (E). All images are 7.5~µm~$\times$~7.5~µm in size. When the surface pressure is increased the nearest neighbor distance is decreased (see SI for further information). In Figure \ref{MainImage}~A, the cores are evenly distributed throughout the entire image and in 2D no preferential orientation of the cores is recognized. Hence, the elliptical core does not influence the ordering or orientation of the microgels. In this height micrograph, the color differs within some of the ellipsoids. When analyzing the height images in more detail (see SI), it becomes clear, that some ellipsoidal cores were not deposited flat onto the substrate but are tilted with respect to the interface. Hence, the thick microgel shell makes it less sensitive to the presence of the core and allows the stabilization of the microgel at the interface with a tilt angle larger than 0~°.\\
Decreasing the surface area, \emph{i.e.}, increasing the surface pressure, (Figure \ref{MainImage}~B-E) leads to an increase in microgel number within one AFM image. Additionally, a favored alignment of the ellipsoidal cores perpendicular to the direction of compression applied with the barriers is observed. The microgels maintain an anisotropic shape due to the elliptical core when the microgel shell is compressed. The ellipsoidal cores rotate preferentially at the interface by revealing the larger soft microgel shell at the sides towards the direction of compression induced by the barriers. Additionally at high surface pressure, some clusters of t-t configuration are observed. This effect is highlighted within the inset of Figure~\ref{MainImage}~E. The quantitative effects of the surface pressure onto the t-t and s-s ordering will be discussed further below. \\
The AFM height images at different stages of the compression isotherm of CS-254 are shown in the second row of Figure~\ref{MainImage} (G-L). The amount of microgels on the observed area at low surface pressure is higher compared to CS-356. Additionally, the ellipsoidal cores show an approximately constant height value in the height images indicating a tilt angle close to 0~°(see SI). From this, we conclude that to reduce the surface tension the ellipsoids are flat on the decane-water interface.\\
By increasing the surface pressure with the help of the barriers, the nearest neighbor distance of the microgels is reduced (see SI). The microgels seem to exhibit behavior commonly observed for anisotropic particles:\cite{Botto2012} Small microgel t-t clusters are observed when the surface pressure is increased (Figure~\ref{MainImage}~I). The appearance of these clusters indicates the existence of attractive forces due to capillary interactions. Such capillary interactions could not be observed within the simulations as the microgels are much smaller and capillary interactions are no significant force within the length scale which are examined with these simulations.\\
Clustering is frequently observed for anisotropic particles indicating the existence of capillary interactions. However, for spherical particles clustering occurs only in specific cases for example for larger\cite{Scheidegger2017} or heavy\cite{Rauh2017} particles. Due to their weight, large particles deform the interface and capillary interaction becomes relevant.\cite{Cohin2013} For CS-254, the compression of the monolayer induces attractive capillary interactions during the isostructural transition. As the weight of every single microgel is constant when compressed at the interface and additionally the microgels do not increase in size, we can conclude that the attractive capillary interactions are induced by a change in anisotropy (aspect ratio) and variation in the degree of immersion of the solid core into the different liquids. This indicates that the compression of the monolayer increases the anisotropy of the microgels at the interface.\\
Here, the capillary interactions lead to clusters with t-t ordering. As discussed above, this is explained in the literature\cite{Botto2012} by shape-dependent capillary interactions. The t-t clustering occurs from far-field capillary interaction for ellipsoids, whereas s-s clustering occurs for near-field capillary interactions.\cite{Botto2012} Hence, the shell is so thick that the elliptical cores are too far from each other to form stable s-s ordering.\\
Some clusters increase in size when the surface pressure is increased to even higher values (Figure~\ref{MainImage}~J), meaning that more microgels contribute to one cluster. Nevertheless, an orientation in one direction as for CS-356 is not observed. Hence, the driving force to rotate the clusters within the direction of compression is not large enough. The microgel shell is compressed further when the surface pressure is increased even more. As a result, the ellipsoids are in close contact and no prominent alignment or ordering is present (Figure~\ref{MainImage}~K). In conclusion with the further compression of the monolayer, the ordering which was observed for less occupied area fraction disappears.\\ 
The AFM height images of CS-165 differ from the other two microgels which is in agreement with the completely different compression isotherm (Figure~\ref{MainImage}~M-R). At low surface pressure, the ellipsoids are colored evenly. Hence, they have a tilt angle of 0~°. Furthermore, the ellipsoids are not separated equally throughout the investigated area but clustering occurs already in the diluted regime (Figure~\ref{MainImage}~M). The clustering is explained by the anisotropic shape of the microgels at the interface. In the case of CS-165, the shell is thinner and additionally the swelling degree smaller compared to the other two microgels. As a result, when the microgel is attached to the interface, the shell spreads less and maintains its anisotropic shape. This anisotropic shape induces attractive forces due to the capillary interactions, as discussed previously.\cite{Botto2012} Nevertheless, no apparent ordering or alignment is visible.\\
Similar to the other two microgels, the nearest neighbor distance decreases when the available surface area is decreased (see SI). Consequently, more microgels are investigated in one image. The amount of microgels located inside the 7.5~µm~$\times$~7.5~µm area is larger as compared to CS-356 and CS-254. The available surface area is reduced and the microgels come closer to one another but additionally keep the inhomogenous distribution/ the clustering (Figure~\ref{MainImage}~N). Hence, the attractive interactions are unaffected by the reduced surface area.\\
At the point where the microgels have to pack closely (Figure~\ref{MainImage}~O), they form short chains of s-s ordering. Consequently, the microgel shell is not too thick to prevent near-field capillary interactions\cite{Botto2012} and the s-s interaction is preferred. Some chains align additionally in t-t leading to small nematic ordering in smaller parts of the monolayer. At region III of the compression isotherm, these chains are unaffected by the direction of the compression and, in conclusion, not aligned in a particular orientation (Figure~\ref{MainImage}~O). In region IV the monolayer is compressed even further and, as a result, the empty areas within the monolayer shrink and a nearly homogeneous and ordered monolayer of anisotropically shaped microgels is formed (Figure~\ref{MainImage}~P). The large s-s microgel chains maintain in region V and only react with alignment to the direction of the compression (Figure~\ref{MainImage}~Q). It has to be noted that the number of particles from region IV to region V does not change drastically from 721$\pm$9 for the six images in region IV as compared to 741$\pm$17 for the six images in region V. Hence, no huge increase in the number of microgels between region IV and V can be observed in the AFM height images. Luo et al. showed a similar clustering of ellipsoidal particles with an aspect ratio of 2 in their simulations. These clusters were a result of the capillary interactions considered in their study.\cite{Luo2019} For CS-165 with an aspect ratio of 1.8 clustering occurs and clusters are formed which are similar to the clusters observed in these simulations. Hence, the clustering of CS-165 is a result of the capillary interactions appearing for anisotropic particles. The main difference between the simulation of Luo et al.\cite{Luo2019} and the experiment in this study is that the clusters in simulations were not compressed with external barriers within one direction and, as a result, no prominent orientation of these clusters was observed. Furthermore, the anisotropic core-shell microgels within this study are not perfectly monodisperse like the particles within the simulation study leading to a lower long range ordering in the experimental study. \\
The images F, L and R in Figure~\ref{MainImage} show the respective core-shell microgels after expansion by opening the barriers. The AFM images show monolayers similar to the monolayers observed at low compression. The initial monolayer at low surface pressure is obtained after the expansion of the monolayer. It is found that the additional clustering in CS-254 and CS-165 is not stable and the microgel monolayers are reversible at low surface pressure. This leads us to the conclusion that the capillary interactions induced by the compression of the monolayer and, hence, the compression of the microgel shells are weak compared to the initial capillary interactions observed for CS-165.\\
The influence and appearance of the capillary interactions are further visualized within Figure~\ref{NND_plot} (left). The most frequent value within the histograms of the nearest neighbor distances (NND$_{frequent}$) (see SI) is plotted against the number of cores (N) within one AFM image. The decrease in NND$_{frequent}$ with increasing N is pronounced and unhindered for CS-356 (green full pointing up triangles). In contrast to that, the dependency of NND$_{frequent}$ for CS-254 is more complicated (blue empty pointing down triangles). First, a similar decrease observed for CS-356 is visible. In this section, the hybrid microgels are not influenced by capillary interactions induced by the anisotropic core. Then, a constant value for NND$_{frequent}$ is maintained for increasing N. This could be an indication for the capillary interactions induced by the anisotropic core as only more microgels align within these t-t clusters with a similar nearest neighbor distance. In the last section, NND$_{frequent}$ decreases again. This is an indication for the annulment of the capillary interactions which results into the not ordered monolayer with smaller nearest neighbor distances between the cores as observed in region V for CS-254.\\
The black trend line shows the numerical connection between NND$_{frequent}$ and N for an isotropic compression of the microgels. The factor of 7.52 is calculated with the NND$_{frequent}$ and N at the contact point. The deviation from this trend is visible especially for CS-254 within the intermediate regime in which the capillary interactions appear.\\
The dependence of NND$_{frequent}$ to N for CS-165 (orange full pointing down triangles) differs from the other two core-shell microgels as the trend line does not represent the data points at all. The data points show no great first decrease but a rather constant value for NND$_{frequent}$. Hence for this microgel, the attractive capillary interactions reduce the influence of the nearest neighbor distance on the compression drastically.\\
Capillary interactions have a significant impact on the ordering behavior of the core-shell microgels within this study. Hence, the sketch in Figure~\ref{NND_plot} (right) visualizes the stiffness and length, and width of the particles, and the consequences for the presence of capillary interactions. For spherical particles, an increase in size and stiffness leads to stronger capillary interactions. For the anisotropic particles prolate and oblate ellipsoid, capillary interactions are more frequently observed for high anisotropy. Furthermore, Figure~\ref{NND_plot} (right) shows an estimate of the position of the three core-shell microgels in the diluted state within this dependence. The microgels are visualized as circles in their respective colors in their initial state. All three microgels have a larger length compared to the width and, hence, are located within the prolate region of this diagram. The sketch indicates the existence of capillary interactions for CS-165 (orange) in contrast to CS-356 (green) and CS-254 (blue) where no capillary interactions are present. Due to the low anisotropy and the soft shell, CS-356 and CS-254 show no capillary interactions whereas CS-156 has higher anisotropy and stiffness resulting in capillary interactions.

\begin{figure}[ht]
\centering
    \includegraphics[width=1\linewidth, trim={0cm 0cm 0cm 0cm},clip]{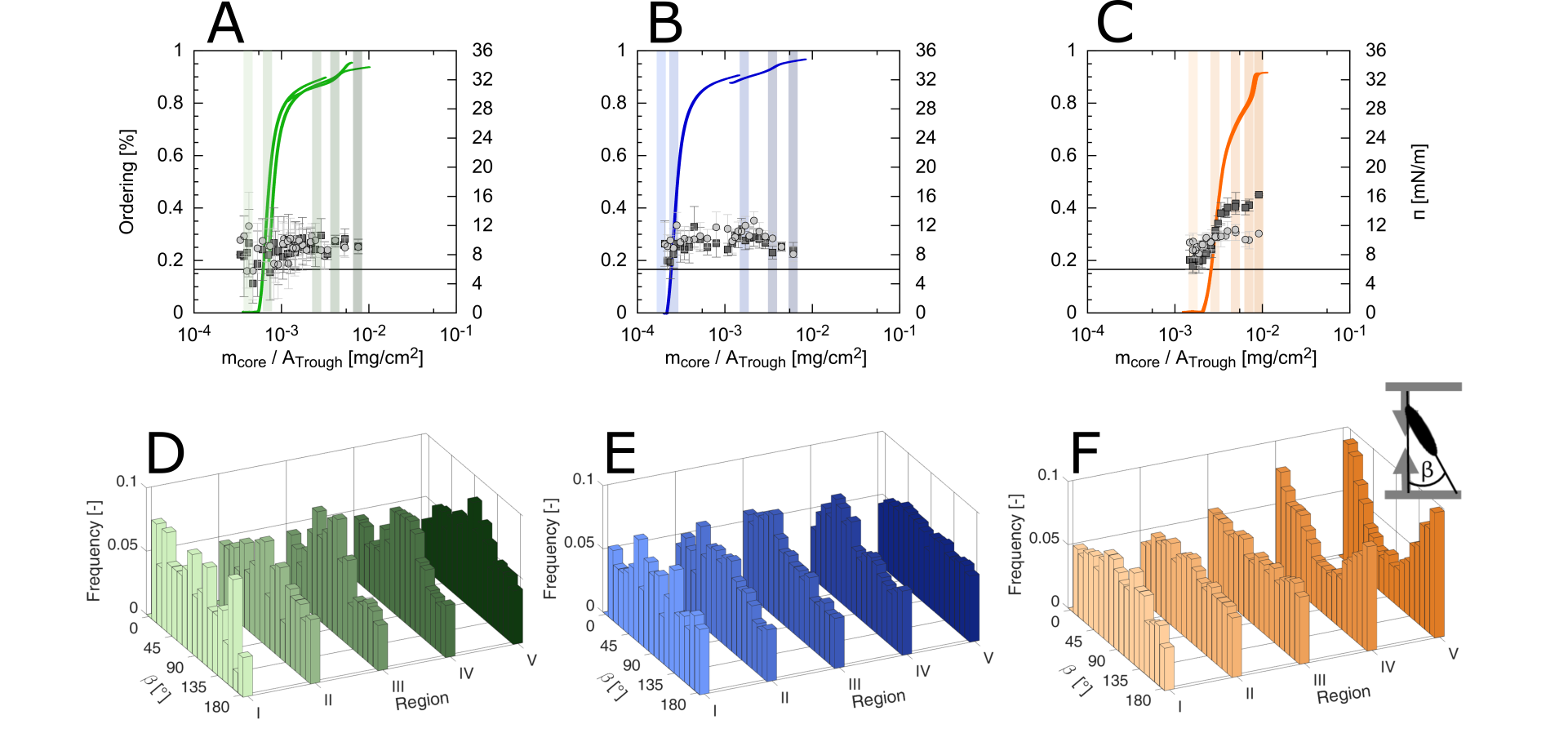}
	\caption{Compression isotherms of CS-356 (A), CS-254 (B) and CS-165 (C) are shown with their quantitative analysis. For each microgel the t-t (circle) and s-s (square) ordering and their depending on the mass of the core divided by the trough area is shown. Additionally, for each compression region a histogram of the angle ($\beta$) distribution reveals the connection between the direction of the compression to the different regions of the compression isotherm (D-F). The schematic illustrates the definition of $\beta$.  The data points taken for the histogram are marked accordingly.}
	\label{OrderingImage}
\end{figure}

Clearly, the AFM height images show a variety of monolayer structures that can be obtained with anisotropic hybrid core-shell microgels. The different shell sizes result into monolayers with different nearest neighbor distances and different arrangements of the microgels and, additionally, a different overall orientation influenced by the direction of the compression. A quantitative analysis gives more information about the ordering and enables to assign the arrangements to positions on the compression isotherms. We investigated the percentage of microgels that are in t-t or s-s contact with their nearest neighbors. The procedure is done with a MATLAB script. First, the script indicates the number of microgels on one image. Then, the script goes through all microgels and checks the position and alignment of all neighboring microgels to search for neighboring microgels with t-t and/or s-s contact. If such a contact is discovered, the centered microgel will be counted as ordered either as t-t, s-s or both. As a last step, the overall number of microgels is taken into account to calculate the percentage of t-t and s-s ordering in one image. These values are plotted within Figure~\ref{OrderingImage} as grey circles (t-t ordering) and dark grey squares (s-s ordering) for all three microgels. Detailed information about the image analysis is provided in the experimental section. The compression isotherms of the deposition of CS-356 are shifted compared to the compression isotherm in Figure~\ref{IsothermImage}. A higher stock solution was used for the experiment in Figure~\ref{OrderingImage}~A compared to the green compression isotherm in Figure~\ref{IsothermImage}. As a result, the probability of microgels lost to the water phase and the error within the normalization in the x-axis increases, leading to a shift to higher values within the x-axis for CS-356.

The values for t-t and s-s ordering of CS-356 (Figure~\ref{OrderingImage}~A) and CS-254 (Figure~\ref{OrderingImage}~B) are approximately constant during the compression. Additionally, no clear preference between s-s and t-t ordering is observed for each of the microgels as the values are scattering in a similar range. Especially the values for CS-356 show large error-bars hindering a further analysis of the data. A reason for the high error-bars could be the low number of hybrid microgels located within the investigated area. Comparing these results to Figure~\ref{OrderingImage}~C indicates that in contrast to the other two hybrid microgels, CS-165 shows the influence of the available trough area onto the s-s and t-t ordering. In region I, the values for the t-t ordering are slightly larger compared to the values for the s-s ordering. This indicates a preferred t-t ordering. This leads to the conclusion that the microgel shell induces far field capillary interactions. When the microgels feel each other the surface pressure increases. Additionally, the s-s ordering increases and additionally the percentage of t-t ordering increases only slightly, leading to a higher amount of s-s compared to t-t ordering at the end of region II. Obviously, the anisotropic microgels adjust in the stable configuration for ellipsoids as short s-s chains. The favoring of the s-s configuration is additionally visible for region III, IV and V. That the initial preferred ordering is t-t, proves that choosing a larger angle for s-s ordering compared to the angle chosen for t-t ordering does not influence the possibility to show the existing preference in t-t ordering. Furthermore, the robustness of the increase in ordering is demonstrated as both ordering types increase even if the nearest neighbor distances decreases. This decrease in nearest neighbor distances lead to a smaller probability to find ordering as the area in which the center of the microgel has to be located decreases (see image analysis for further details).\\
The evaluation regarding the amount of t-t and s-s ordering considers only the nearest neighbors of each microgel. Figure~\ref{OrderingImage}~D-F shows the angle ($\beta$) of all microgels as histograms to visualize the alignment of the monolayer. $\beta$ is defined as the angle between the microgels long axis and the direction of the compression (see Figure~\ref{OrderingImage} sketch on the right). For each microgel one histogram for each of the five different regions of the compression isotherm is shown to demonstrate the evolution of the distribution of $\beta$ in respect to the compression of the monolayer. The histograms (Figure~\ref{OrderingImage}~D-F) are linked to the compression isotherms (Figure~\ref{OrderingImage}~A-C) by the vertical boxes using the similar color coding as in the histograms. The histograms of $\beta$ change within the progression of the compression isotherms for all three microgels. Hence, the microgels react to the compression of the monolayer by adjusting their overall alignment.\\
The microgel with the largest shell (CS-356) shows randomly distributed $\beta$ for region I and II (Figure~\ref{OrderingImage}~D). These random peaks could be a result of the low number of microgels resulting in poor statistics within these regions. Regions III, IV and V indicate a higher probability of $\beta$ around 90~° for the ellipsoidal microgel cores. Additional to the quantitative analysis, the qualitative analysis shows a preferred ordering of the microgel cores of CS-356 perpendicular to the direction of compression at higher surface pressures.\\
The microgel with the intermediate shell size (CS-254) shows a random distribution for $\beta$ in region I and II. No overall orientation is observed for the microgels when they are without contact or only their shells are in contact. In region III, the microgel cores show a small tendency towards $\beta$ = 45~°. This means within the coexistence regime, the direction of the compression influences the alignment of the microgel cores. This effect is even more pronounced in region IV. $\beta$ of 45~° means that the ellipsoidal cores within the microgels are neither orientated with the direction of compression nor perpendicular to it, but some ordering in between is preferred. This preferential alignment is lost in region V where a homogeneous distribution of $\beta$ is observed. As the shell is thinner compared to CS-356 the urge to align the thick shell on the sides towards the direction of compression is less pronounced and moreover completely lost in region V. The loss of the preferred alignment corresponds to a small decrease in the values of the s-s and t-t ordering observed in Figure~\ref{OrderingImage}~B.\\
In computer simulations the microgel with the largest (CS-L) and intermediate (CS-M) shell shows the same tendency. The peak on the angle distribution histogram around $\beta$ = 90~° and $\beta$ = 45~° for the (CS-356) and (CS-254) samples qualitatively corresponds to the peak of the scalar order parameter. Similar to the simulation, the maximum peak values are located in region III and become less pronounced with compression and, in case of (CS-254), is even completely lost in region V.\\
CS-165 shows the largest changes within the histograms. The regions I, II and III indicate a homogeneous distribution for $\beta$ of the microgels. In contrast to the other two microgels, the effect of the direction of compression does not influence the orientation of the microgels within region III. Only when the microgels reach region IV, a preferential alignment of 0~° and 180~° is observed. The histogram visualizes that the microgels align in direction of the compression of the barriers. The s-s chains formed in the previous region are pushed perpendicular to the compression and, as a result, the microgel cores align parallel to it. This preferential alignment is even more pronounced in region V. In the simulations, the microgels with the smallest shell (CS-S) show a shift of the scalar order peak to lower $Ap$ values (region IV and V) similar to the experiments. The CS-165 microgels are highly influenced by the direction of the compression when the monolayer is compressed to the point where an almost fully covered interface and a highly ordered microgel monolayer is reached.

\section{Conclusions}
\begin{figure}[ht]
\centering
\includegraphics[width=0.5\linewidth, trim={0cm 0cm 0cm 0cm},clip]{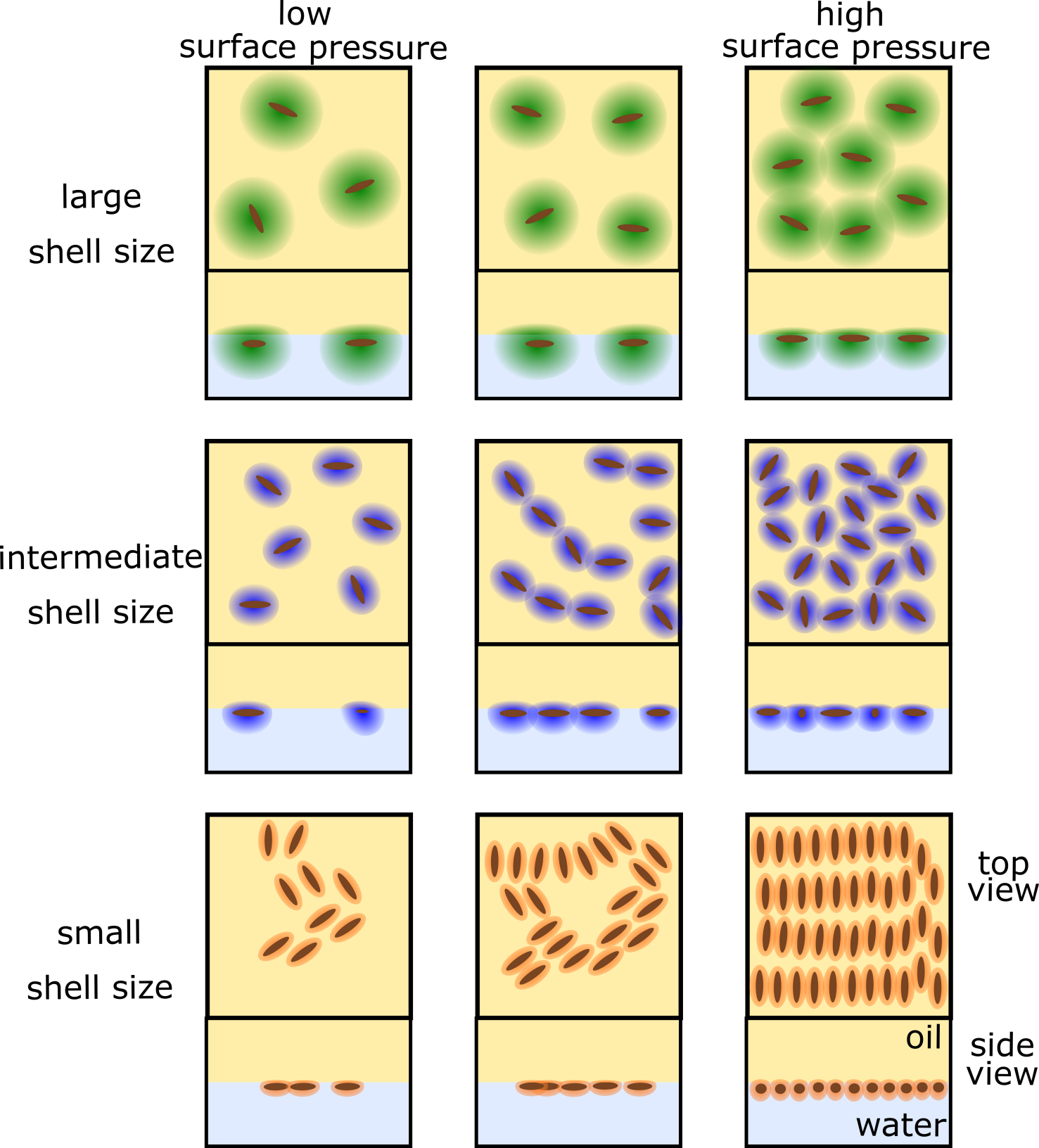}
	\caption{Sketch of the ordering behavior of anisotropic hybrid microgels. Different microgel shell sizes allow either a tip-to-tip (left), side-to-side(right) ordering or no ordering at all (middle) when the microgel monolayers are compressed at a decane-water interface. The direction of compression is vertical for the top view images identical to the direction of compression in Figure~\ref{MainImage}.}
	\label{result}
\end{figure}

In conclusion, we showed the rich two dimensional phase behavior at the decane-water interface of anisotropic hybrid core-shell microgels having the same core but different microgel shells. A combination of the Langmuir-Blodgett technique with ex situ AFM imaging as well as DPD simulations revealed the influence of different microgel shells on the ordering and orientation of the hybrid core-shell microgels. Soft microgels with a thick shell strongly spread at the interface, which is much reduced for microgels with a harder and thinner shell. As a consequence, the latter exhibit clusters in the diluted state, in contrast to the larger microgels which do not cluster. The elliptical hematite-core facilitates the anisotropic shape of the different microgels and the capillary interactions which result from the shape. We show that by adjusting the shell thickness and stiffness, we tune capillary interactions of these anisotropic microgels.\\
Even more interesting characteristics of these anisotropic microgels are observed at higher compression states in the Langmuir-Blodgett trough. Their anisotropy makes them sensitive to the direction of compression in the Langmuir-Blogett trough. The not compressible anisotropic core rotates to minimize the emerging stress. The combination of the anisotropic shape, the softness of the shell and the compression within the Langmuir-Blodgett trough reveal characteristic monolayers with different ordering and orientation of the microgels. As a result, the adjustment of the stiffness and thickness of the microgel shell leads to different effects at the interface. This results into completely different monolayers for the three core-shell microgels. Figure~\ref{result} illustrates those effects in dependence of the shell size and the surface pressure. The schematics demonstrate how the microgels act when compressed at the interface. For this, the top and side view for the three different core-shell microgels at low surface pressure, intermediate surface pressure and high surface pressure are presented.\\

\noindent With CS-356, we synthesized a microgel where the shell is thick enough to erase the capillary interactions between the cores in the dilute phase and an effect caused by their anisotropic shape only reveals due to the compression of the shell with the barriers. When the interface is compressed, the elliptical cores rotate in such a way that most part of the compressible microgel shell aligns in the direction of the compression. As a result, the elliptical cores align perpendicular to the direction of compression. This result is confirmed by the simulation of the monolayer of CS-L microgels with thickest shell.\\
An intermediate behavior is observed for the intermediate shell thickness (CS-254): The microgel shell erases the capillary interactions  at low surface pressure. While compressing, a regime is obtained where capillary interactions are enhanced as the shell is not thick enough to prevent it and t-t clustering occurs. Nevertheless, these t-t clusters are not stable enough to resist further compression and as a result, the order decreases upon further compression. The simulations qualitatively reproduce this results. However, the effect of capillary forces is difficult to assess probably due to the small size of the microgel, as well as their small number in the simulation box \\
The microgels with the smallest and additionally hardest shell (CS-165) show self-assembly due to capillary interaction induced by the anisotropic shape even in the dilute state. Especially the s-s ordering is triggered during the compression. Compared to the t-t clusters of CS-254, these s-s clusters are stable enough and exist during the compression and react collectively by re-orientating complete clusters.\\
Our results show the importance to consider the softness additionally to the shape when anisotropic particles and their ordering are investigated. Furthermore, the combination of a soft microgel shell with an elliptical hard core provides new opportunities for example in nanolithography\cite{Rey2016_NL}. By adjusting the shell thickness, one can build ordered monolayers of microgels with a specific orientation. This can be used for the fabrication of nanowires. For future experiments, the characteristic shapes of the compression isotherms, especially in the case of CS-165, can be used to gain information about the present monolayer without the need of additional imaging techniques. The information about a more isotropic or more anisotropic behavior at the interface is obtained by investigating the shape of the compression isotherm. Afterwards, a specific monolayer is reached by applying a specific surface pressure at the desired compression state and no further ex situ AFM analysis is needed. With this study, we show effects of the combination of softness and anisotropy onto the attractive capillary interactions. Future studies with larger aspect ratios for the core can be envisioned to investigate further limits of these shape dependent capillary interactions. Furthermore, Grillo et al.\cite{Grillo2020} showed a procedure in which complex microstructures of spherical microgels are obtained by using a two step deposition process. The use of anisotropic microgels in this type of experiment may provide access to further complex microstructures beyond the limit of isotropic particles.

\begin{acknowledgement}
We thank Nicole Terefenko for the synthesis of microgel CS-165. Financial support from the SFB 985 “Functional Microgels and Microgel Systems” of Deutsche Forschungsgemeinschaft and the Russian Foundation for Basic Research (project nos. 19-03-00472 and 20-33-70242) is greatly acknowledged. 
\end{acknowledgement}

\begin{suppinfo}
The following files are available free of charge.
\begin{itemize}
  \item Experimental data for this publication is available per request at the persistent identifier: https://hdl.handle.net/21.11102/b71bcdb5-c386-11eb-afb2-e41f1366df48
  \item Supporting information: Dynamic Light Scattering, Thermogravimetric Analysis, Regions within the Compression Isotherms, Nearest Neighbor Distance, Atomic Force Microscopy Height Profiles, Experimental Section of Dynamic Light Scattering and Thermogravimetric Analysis, Computer Simulations Parameters and Calculations, Simulations of microgel characterization in bulk, Effect of the box and influence of the method of preparation of compressed systems
\end{itemize}

\end{suppinfo}

 \bibliography{main} 
 
 \newpage
 \begin{figure}[ht]
\centering
\includegraphics[width=8.25cm, trim={0cm 0cm 0cm 0cm},clip]{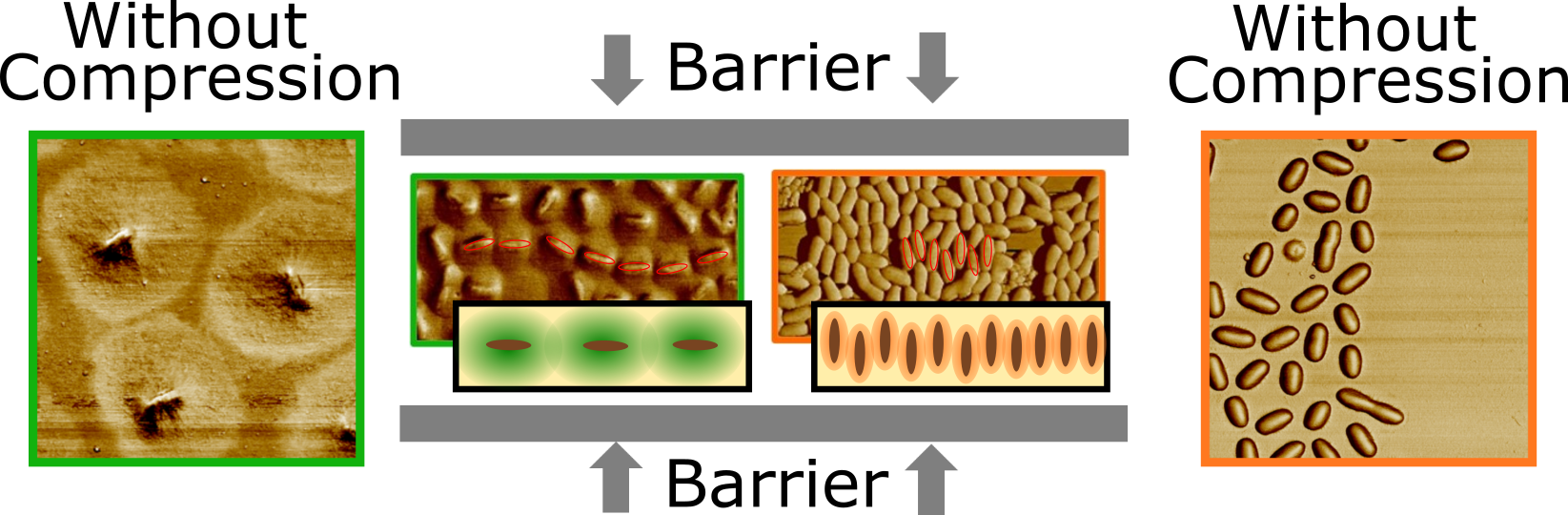}
	\caption{TOC Graphic}
\end{figure}

\end{document}


\section{Dynamic Light Scattering}
\begin{figure}[ht]
\centering
    \includegraphics[width=0.5\linewidth, trim={0cm 0cm 0cm 0cm},clip]{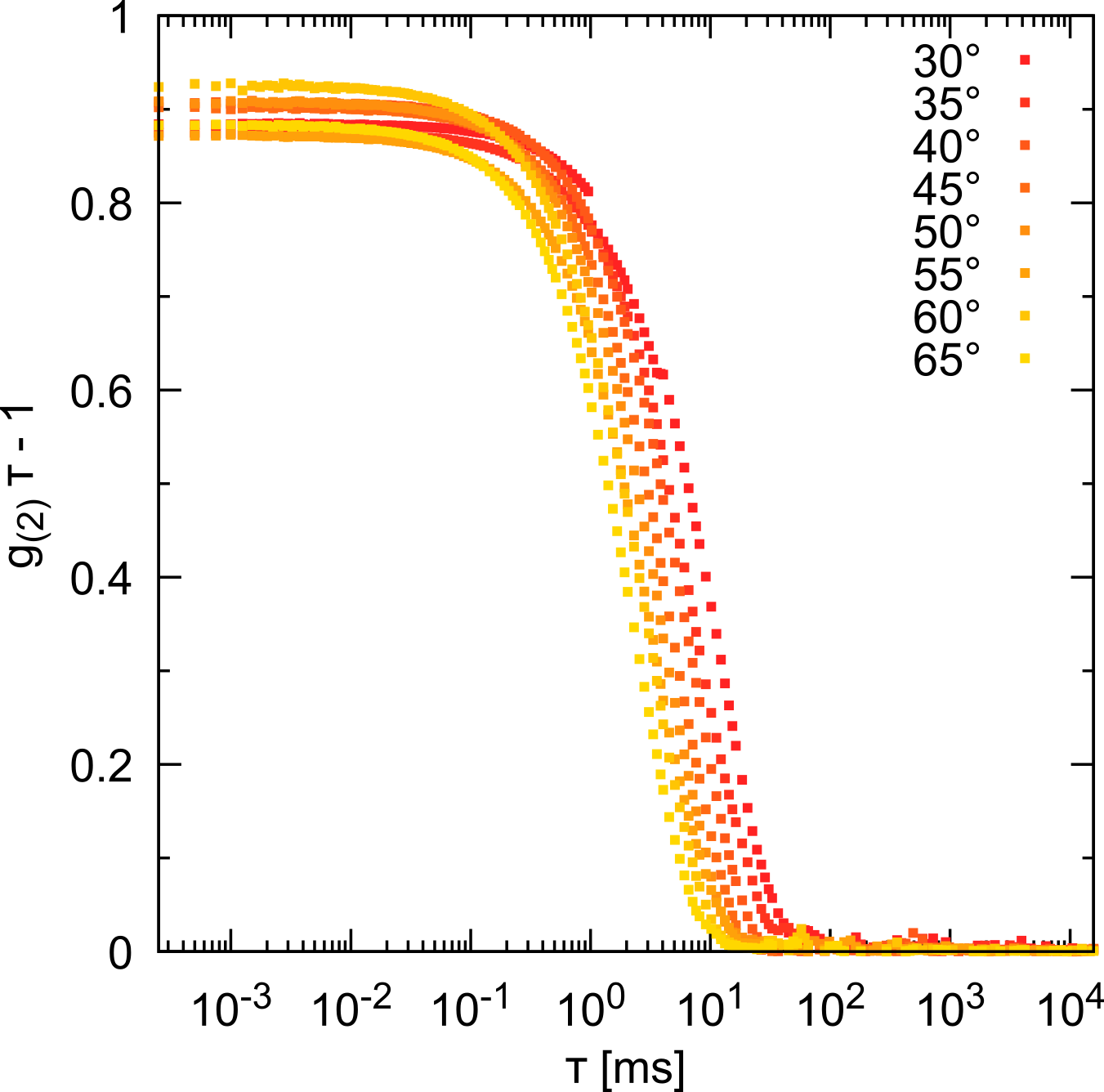}	
	\caption{Correlation functions obtained from DLS measurements of CS-356 at 10~°C at various angles.}
	\label{Correlation_DLS}
\end{figure}

Dynamic light scattering (DLS) measurements are used to determine diffusion coefficients of submicrometer-sized particles. Due to the Brownian motion of such particles a time-dependent fluctuation of the scattered intensity is observed. If the auto-correlation function obtained from DLS measurements shows a one step decay (see Figure~\ref{Correlation_DLS}), the diffusion coefficient is obtained via the cumulant fit. The second-order cumulant was plotted against $q^{2}$ and the diffusion coefficient was calculated by the linear regression.

The hydrodynamic radius $R_{H}$ is calculated via the Stokes-Einstein equation

\begin{equation}
     R_{H} = \frac{k_{B}T}{6\pi\eta D}.
     \end{equation}

\begin{figure}[ht]
\centering
    \includegraphics[width=0.5\linewidth, trim={0cm 0cm 0cm 0cm},clip]{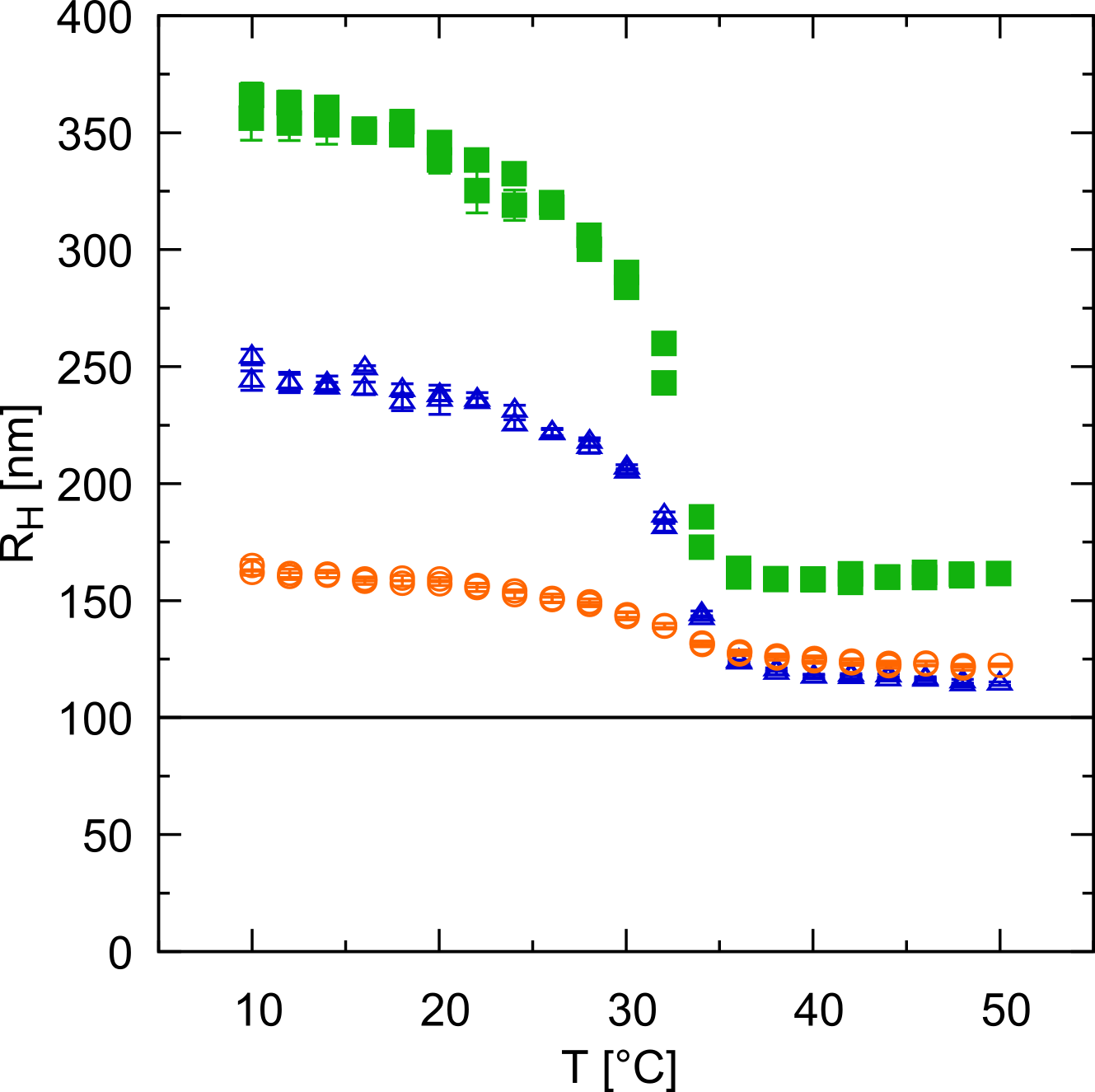}	
	\caption{Hydrodynamic radii dependent on the temperature for CS-356 (green), CS-254 (blue) and CS-165 (orange). All three microgels show a reversible deswelling with increasing temperature.}
	\label{RH_DLS}
\end{figure}

\noindent For spherical microgels $R_{H}$ is an important characteristic and temperature dependent DLS measurements are important to obtain information about the swelling behavior of such microgels. For anisotropic microgels $R_{H}$ an apparent $R_{H}$ gives information about the characteristics of the collapse and swelling of the microgel. One decay is observed in the auto correlation functions in the case of the microgels in this study. Figure~\ref{Correlation_DLS} shows the raw data from a set of all angles at 10~°C of CS-356. We will use the hydrodynamic radius to obtain information about the overall size differences of the microgels and to identify their temperature responsiveness.\\
Figure~\ref{RH_DLS} shows the combined DLS results for the three microgels. For each microgel the $R_{H}$ of a heating and cooling curve from 10 to 50~°C are shown. Furthermore, a black horizontal line indicates the hydrodynamic radius of the elliptical hematite-silica core which is not temperature responsive.   

The green data points in Figure~\ref{RH_DLS} show the $R_{H}$ of CS-356. The microgel has the largest shell indicated by the highest values for the $R_{H}$. Furthermore, the expected decrease in the radius is observed when the microgel is heated up to 50~°C. The blue data points show a similar temperature responsive behavior of CS-254. Nevertheless, the initial $R_{H}$ at 10~°C is approximately 100~nm smaller compared to CS-356. The last microgel CS-165 (orange data points) has the smallest initial $R_{H}$ at 10°C. Furthermore, the less pronounced decrease in size , nevertheless, still a temperature responsiveness is observed. In the collapsed state at 50~°C, CS-254 and CS-165 have a similar size which is smaller when compared to the size of the collapsed microgel CS-356. In conclusion, the DLS data show that the investigated microgels all have differently sized and differently soft shells and all show a temperature responsive behavior which is typical for microgels synthesized from the monomer NIPAm.

\section{Thermogravimetric Analysis}
\begin{figure}[ht]
\centering
    \includegraphics[width=1\linewidth, trim={0cm 0cm 0cm 0cm},clip]{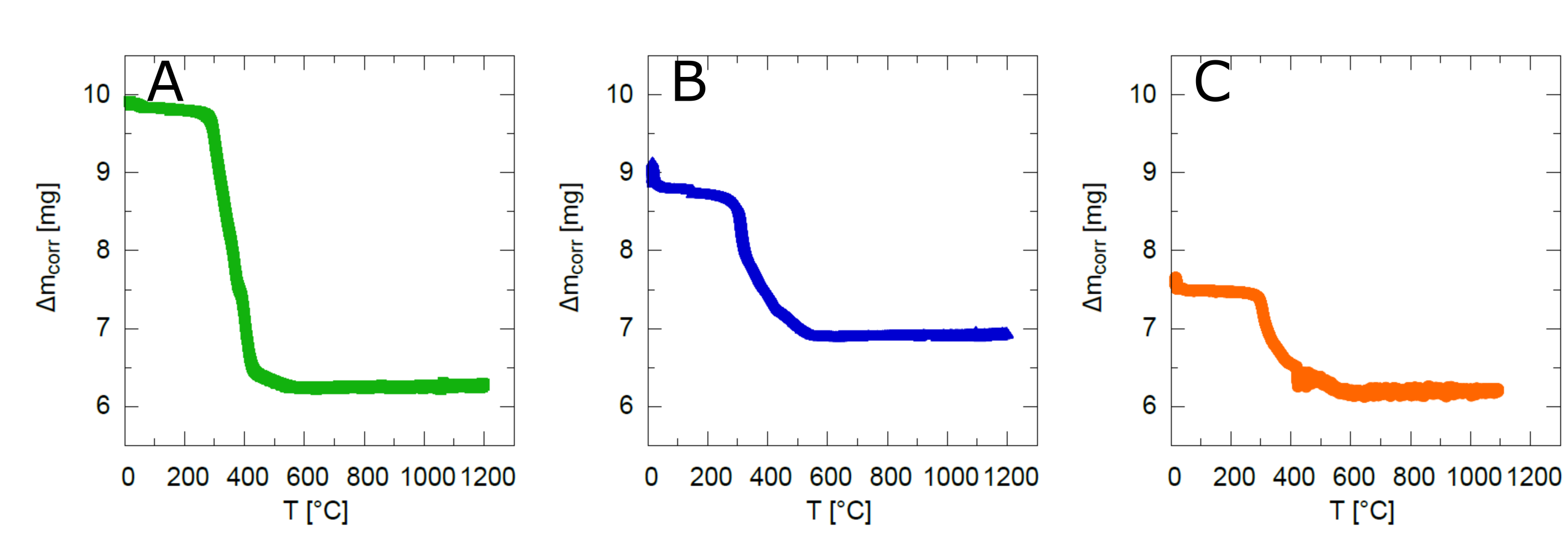}	
	\caption{Mass loss during the thermogravimetric analysis for CS-356 (A), CS-254 (B) and CS-165 (C). All microgels show a mass loss when heated above approximately 300~°C.}
	\label{TGA}
\end{figure}

The thermogravimetric analysis (TGA) is used in this study to quantify the composition of the different anisotropic core-shell microgels. The same elliptical hematite-silica core is used for all three samples and they only differ in their respective shell masses. With TGA the samples are heated to 1200~°C with a heating rate of 1~K$\cdot$min$^{-1}$ in a controlled atmosphere of 20\% oxygen in nitrogen at ambient pressure. The buoyancy corrected mass difference is plotted in dependence of the temperature (Figure~\ref{TGA}).

All three measurements show a similar behavior: After evaporation of solvents and adsorbed water at or below 150~°C, the mass is approximately constant up to 275~°C. A sharp decrease of the mass is visible in each of the three curves upon a further increase of the temperature, caused by the decomposition of each of the microgel shells. In conclusion, the stability of the polymer shell is limited to temperatures up to approx. 275~°C. The steepness and shape of the drop is different for all three microgels. Above 600~°C, no further decomposition can be observed and the mass is constant up to 1200~°C. Nevertheless, for this study only the initial mass before, $\Delta m_{TGA,initial}$, and the final mass after the decrease,  $\Delta m_{TGA, final}$, have been evaluated. With the masses before and after the decrease, the original mass of the polymeric shell, $m_{shell}$, is calculated.

\begin{equation}
     m_{shell} = \Delta m_{TGA,initial} - \Delta m_{TGA, final}
\end{equation}

\noindent The obtained masses are listed in table \ref{table_results}. These masses are subtracted from the original mass of the sample weighted into the aluminum crucible ($m_{core+shell}$) to calculate the mass of the core ($m_{core}$).

\begin{equation}
    m_{core} = m_{core+shell} - m_{shell}
\end{equation}

\noindent All calculated masses and the respective mass percent of core and shell are listed in table \ref{table_results}. The values indicate that CS-356 with the largest shell indicated by the DLS additionally has the heaviest shell with 77.3~\%. In contrast to that, even though CS-254 and CS-165 have a different stiffness and therefore additionally a different swelling behavior, the amount of shell related to the core mass is similar with 60.9~\% for CS-254 and 65.0~\% for CS-165. This is in agreement with the slightly larger shell of CS-165 for the hydrodynamic radii in the collapsed state at 50~°C. In summary, CS-356 has a large and soft microgel shell, CS-254 has a small and soft microgel shell and CS-165 a small and harder microgel shell. This values are approximately similar to the percentages of the monomer (mon$_{i}$) and core (c$_{i}$) weights used for the synthesis. As the core is dispersed in ethanol the mass of the used core is faulty leading to less reliably values.

\begin{table}[ht]
     \centering
    \caption{Comparision of the thermogravimetric analysis (Original mass, shell mass, core mass and the resulting proportions of shell and core respectively) to the initial amount during the synthesis}
     \begin{tabular}{c|ccc|cc|cc}
       \hline
     sample & $m_{core+shell}$ [mg] & $m_{shell}$ [mg] & $m_{core}$ [mg] & shell [\%] & core [\%] & m$_{i}$ [\%] & c$_{i}$ [\%] \\
       \hline
CS-356 & 4.62 & 3.57 & 1.05 & 77.3 & 22.7 & 73 & 27 \\
CS-254 & 3.02 & 1.84 & 1.18 & 60.9 & 39.1 & 62 & 38 \\
CS-165 & 1.97 & 1.28 & 0.69 & 65.0 & 35.0 & 62 & 38 \\
       \hline
     \end{tabular}
     \label{table_results}
     \end{table}
     
\section{Regions within the Compression Isotherms}
\begin{figure}[ht]
\centering
    \includegraphics[width=0.6\linewidth, trim={0cm 0cm 0cm 0cm},clip]{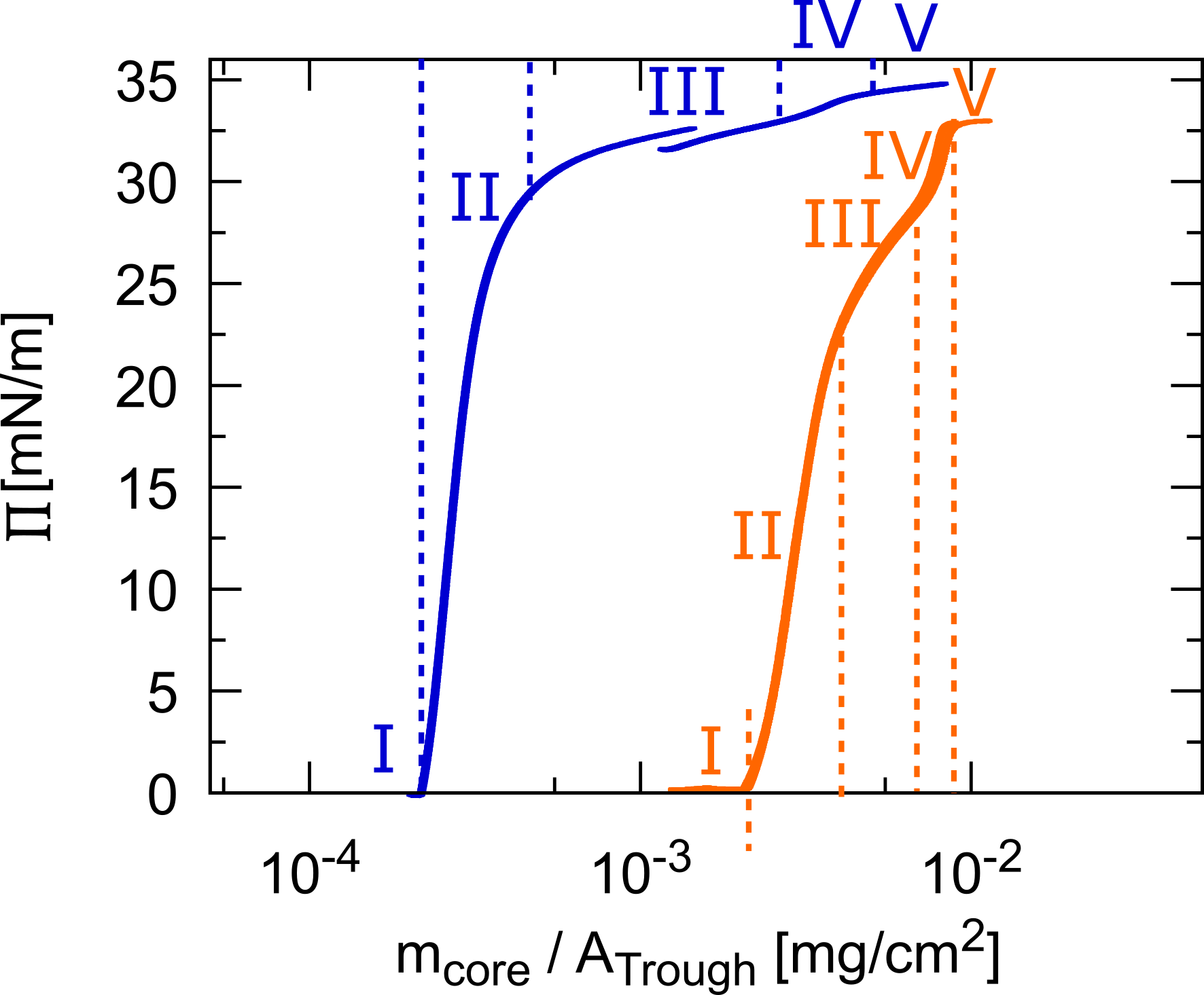}	
	\caption{Compression isotherms of CS-254 (blue) and CS-165 (orange) with the assignment of the five regions within the isotherms.}
	\label{orangeblue}
\end{figure}

Figure~\ref{orangeblue} illustrates the compression isotherms of the microgel with the intermediate shell thickness (CS-254, blue line) and the thin shell (CS-165, orange). Furthermore, the five different regions for both of the anisotropic core-shell microgels are marked with the dotted lines in the respective color. The shape of the different compression isotherms is discussed in the main part.

\section{Nearest Neighbor Distance}
\begin{figure}[ht]
\centering
    \includegraphics[width=1\linewidth, trim={0cm 0cm 0cm 0cm},clip]{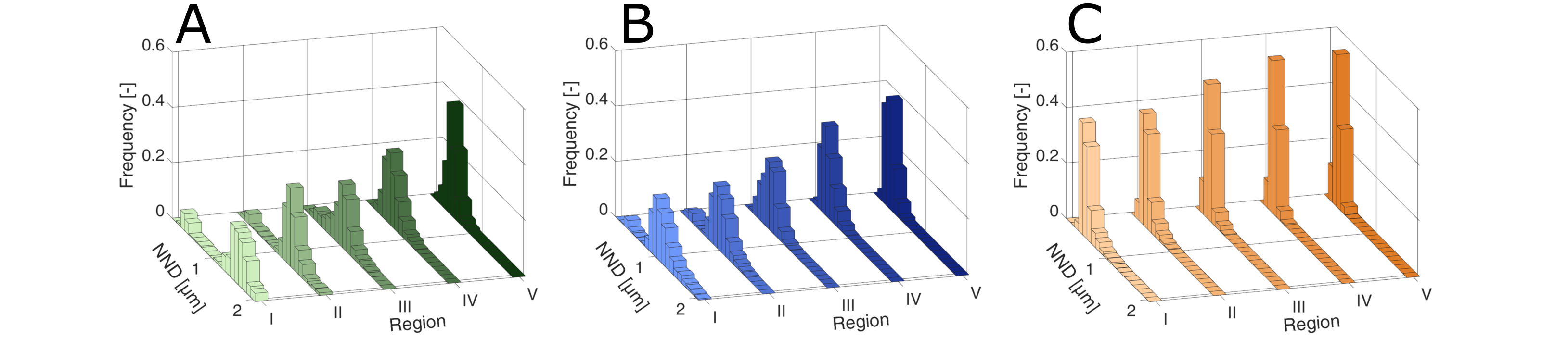}	
	\caption{The nearest neighbor distance histograms of CS-356 (A), CS-254 (B) and CS-165 (C) dependent on the phases according to the compression isotherms.}
	\label{NND}
\end{figure}

For all three microgels in the AFM height images, the nearest neighbor distances were calculated. For each 7.5~µm~x~7.5~µm image the microgel centers were analyzed. By using a Voronoi diagram, the nearest neighbors are detected. For each nearest neighbor, the distance was calculated and plotted within a histogram. In Figure~\ref{NND} the sum of the distances are shown. Hereby, the histograms are based on the same images resulting into the angle histograms in the main part of this work. Hence, each microgel is presented with five nearest neighbor distance (NND) - histograms, one for each of the five phases of the compression isotherm. 

For CS-356 (Figure~\ref{NND}~A), the NND histogram in the dilute phase I shows two groups of nearest neighbor distances. The first one is located between 1~µm and 2~µm. The second one is at lower values of about 300~nm with a lower frequency compared to the first group. As this second group is located around the core length, these values occur properly from microgels with two cores. Before the synthesis is started, the core solution is treated with ultrasonification to disperse single cores within the reaction solution, nevertheless, some cores cluster and serve together as core for one microgel shell. Hence, the peak around 300~nm is present within all phases and gives no information about the change in the distance between two different microgels. To investigate the distance between two microgels, the peak at higher nearest neighbor distances is needed. This peak is shifting within every phase towards smaller values for the NND. In phase I, the histogram shows a broad peak around 1.5~µm. This peak is constantly shifting towards a NND of around 500~nm in the last phase V indicating a compression of the microgel shell and a denser packed microgel monolayer at higher surface pressure.\\
For CS-254 (Figure~\ref{NND}~B), similar histograms compared to CS-356 are obtained. In phase I, a small peak around a NND of 300~nm indicating most properly the small amount of double core microgels which are not further discussed here. The more prominent is the broad peak at higher NND distances around 1~µm. Similar to CS-356, this peak and as a conclusion the NND is shifting to lower values while increasing the surface pressure. Compressing to phase IV leads to a broad peak with a NND maximum at 400~nm, while further increase in the surface pressure (Phase V) the maximum NND of 400~nm remains. The difference between the histograms of these two phases is the width, the broad peak in phase IV is compressed to a thinner peak in phase V.\\
For CS-165 (Figure~\ref{NND}~C), the histograms show much smaller NND compared to the larger microgels in the lower phases. The clustering observed in the AFM images, leads to a NND histogram with only one main peak with a maximum around 400~nm. When the monolayer is compressed and the surface pressure is increased within the different compression isotherm phases, this one peak is only slightly shifted to lower NND to a peak maximum around 300~nm in phase V. Similar to CS-254, the peak width is decreased simultaneously to the shifting of the maximum. \\
To sum up, the NND histograms show an decrease in NND and/or in the width of the NND for all three microgels when the monolayers are compressed. The starting values for the NND are the largest for the CS-356 and the smallest for CS-165, which is in agreement to all other observations. 

\section{Atomic Force Microscopy Height Profiles}
\begin{figure}[ht]
\centering
    \includegraphics[width=0.9\linewidth, trim={0cm 0cm 0cm 0cm},clip]{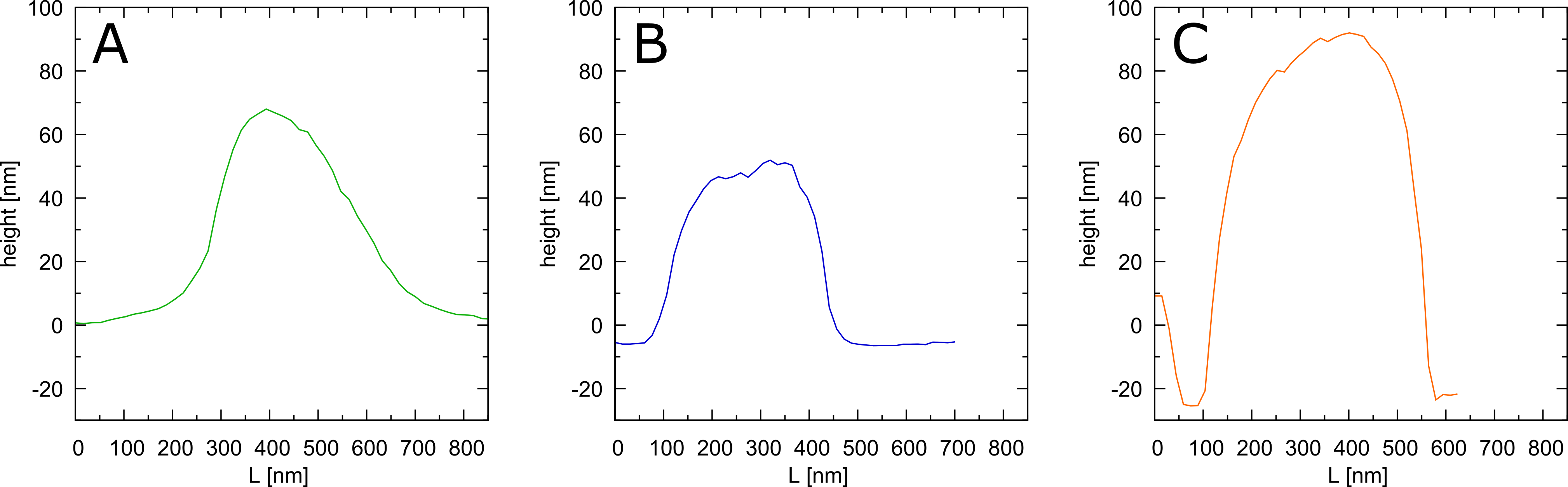}	
	\caption{Examples of a length height profile of one microgel of CS-356 (A), CS-254 (B) and CS-165 (C).}
	\label{AFM_height}
\end{figure}

In the bare AFM height images of the different microgels in diluted state (phase I), the elliptical cores have a different appearance. The images of sample CS-356 show a high polydisperse length and inhomogeneous colored ellipses whereas the images of CS-165 show low polydispersity and homogeneous coloring. As the same batch of cores was used to synthesis the three different anisotropic core-shell microgels, the inhomogeneity exists not due to the core itself. To investigate this effect further, the height profiles sliced within the length of one microgel are investigated. Examples for one microgel in phase I of the compression isotherm of the three core-shell microgels are shown in Figure~\ref{AFM_height}. 

Figure~\ref{AFM_height}~A presents an example of a CS-356 microgel in the dilute regime. The height profile in the length shows a tilt towards the right compared to a centered profile which could be envisioned for microgels with an elliptical core laying flat on the surface. As a result, the core in the here presented microgel is not flat on the surface. As this is just one microgel, is has to be mentioned that similar shapes of the height profile can be observed for other individual microgels in CS-356. Nevertheless, additionally microgels with a flatter profile and hence a flat ellipsoidal core are present. To conclude, for the microgel CS-356 the inhomogeneity of the ellipsoids comes from the different arrangement of the cores on the solid substrate and not the polydispersity of the cores.\\
In contrast to the height profile of one microgel of CS-356, the height profiles of CS-254 (Figure~\ref{AFM_height}~B) and CS-165 (Figure~\ref{AFM_height}~C) show a centered profile indicating microgels and especially their cores laying flat on the solid substrate. Even if the presented profiles present only one example, all investigated profiles of these microgels had a centered profile, which agrees to the monodisperse appearance of the microgels in the images obtained for CS-254 and CS-165.\\
Furthermore, it is possible to compare the different shaped edges of the length profiles of the investigated microgels. The microgel showing the hardest behavior shows the sharpest edges and additionally the highest profile. This indicates that the spreading of the microgel shell is reduced compared to both other microgels which show lower profiles and additionally a smoother decrease at the edges, suggesting a spreading of the microgel shell especially for CS-356.

\section{Experimental Section}
\textbf{Dynamic Light Scattering}\\
The dynamic light scattering measurements were conducted on an ALV 5000 goniometer containing a laser with a wavelength of 633~nm, a digital hardware correlator and two avalanche photodiodes. A toluene bath was used to index-match the surrounding medium. The temperature was adjusted by an external programmable thermostat (Julabo F32). For each microgel, the experiments were performed from 10~°C to 50~°C in 2~°C-steps for the heating curve and in 2~°C steps back to 10~°C for the cooling curve to check for reversibility. For each temperature, the measurements were performed from 30~° to 65~° in 5~° steps and the measurement time for each angle was 60~s. For the measurements, highly diluted samples were used. The cumulant analysis was used to obtain the decay rate for each measurement. The second-order cumulant was plotted against $q^{2}$ and the diffusion coefficient was obtained from the gradient of the linear regression.

\noindent\textbf{Thermogravimetric Analysis}\\
The weight percentage of the core with respect to the shell of the discussed hybrid core-shell microgels was determined by thermogravimetric analysis: A specific amount of the freeze-dried core-shell microgels, $m_{core+shell}$, was weighted into 1300~µL alumina crucibles and placed in a SETSYS 16/18 thermobalance (SETARAM, Paris, France). Under a constant gas flow of 20\% oxygen in nitrogen (both N5.0 purity) at ambient pressure ,the samples were first kept at room temperature for two hours to thoroughly exchange the atmosphere, afterwards heated to 150~°C with a heating rate of 1~K$\cdot$min$^{-1}$ and kept at this temperature for three hours to ensure the evaporation of solvents and adsorbed water. With a constant heating rate of 1~K$\cdot$min$^{-1}$ the samples were subsequently heated to 1200~°C. The decomposition of the polymeric shell was observed between temperatures of 275 $\leq$ T /°C $\leq$ 600 for all samples by a significant decrease of the buoyancy corrected mass.

\section{Computer Simulations}
\textbf{Parameters}\\
The parameter $a_{ij}$ is defined according to the Flory–Huggins parameter, $\chi_{ij}: a_{ij} \approx a_{ii} + 3.27\chi_{ij}$, where $i,j \in \{W,O,M,P)$. First, we set the total number density of the system as $\rho = 3r_{c}^{-3}$ to reproduce the isothermal water compressibility\textsuperscript{S1} $a_{ii}$ = 25. The $\chi_{ij}$ parameters for oil-water and oil–microgel interactions can be calculated with the Hansen solubility parameter\textsuperscript{S2}

\begin{equation}
    \chi_{ij} = \frac{\alpha\nu_{ref}}{k_{B}T}\lbrack (\delta_{i}^{d} - \delta_{j}^{d})^{2} + 0.25(\delta_{i}^{p} - \delta_{j}^{p})^{2} + 0.25(\delta_{i}^{h} - \delta_{j}^{h})^{2}\rbrack,
    	\label{simeq8}
\end{equation}
where $\alpha$ is a numerical coefficient usually taken as 0.6\textsuperscript{S3} $\nu_{ref}$ is the reference average volume of $i^{th}$ and $j^{th}$ beads (the volume of the single DPD bead), and $\delta_{i}^{d}$, $\delta_{i}^{p}$ and $\delta_{i}^{h}$  are the dispersion, polar and hydrogen bonding Hansen solubility parameters, respectively, which were obtained from calculations\textsuperscript{S4} or the existing data\textsuperscript{S2,S5} (Table \ref{tab5Sim}).\\
To define the $\nu_{ref}$, we follow the work of Groot and Rabone\textsuperscript{S6} and define a water mapping number, $Nm$, so that each water bead corresponds, on average, to $Nm$ water molecules. Hence, the characteristic mass of this type of beads will be multiple to 18~Da, and the molecular volume, $\nu_{m}$, will be approximately multiple of $\sim30~$\r{A}$^{3}$. By selecting $Nm$ = 9, the beads should have a mass of around 162~Da and occupy a volume of 270~\r{A}$^{3}$. pNIPAm monomer has a mass of 113~Da and occupy a volume of 180~\r{A}$^{3}$. Thus 1.5 monomers of NIPAm are in the bead. Decane molecule has a mass of 169~Da and occupy a volume of 233 ~\r{A}$^{3}$.\textsuperscript{S7} Therefore, 1 decane molecule is in the bead. We can finally determine the characteristic length, $r_{c}$. By using the mapping number\textsuperscript{S8} $\rho N_{m}\nu_{m} = 1$, we obtain $r_{c}$= 0.93~nm.\\
Knowing the characteristic length, mass and energy, the time scale for the DPD simulation is calculated as $\tau = \sqrt{mr_{c}^{2}/k_{B}T}$ $=$ 7.53~ps.\\
By using equation~\ref{simeq8}, the repulsion parameters for decane–water and microgel–decane pairs could be estimated as $a_{WO}$ = 90.13 and $a_{MO}$ = 30.38, respectively. In the case of pNIPAm-water interaction, we used the approach proposed by Balazs et al. for simulating thermoresponsive gels\textsuperscript{S9} and later adapted for microgels\textsuperscript{S10}. In turn, at T$\sim$20°C, we obtain $a_{WM}$ = 25.6, which corresponds to a good solvent with $\chi_{ij}$= 0.18. All the parameters are presented in Table~\ref{tab6Sim}. All simulations were performed using the open-source software LAMMPS\textsuperscript{S11}.

\begin{table}[H]
     \centering
    \caption{Sizes of the microgel with nanoparticle in a swollen state at T=20°C. }
     \begin{tabular}{ccccccc}
       \hline
& \multicolumn{3}{c}{experiment\textsuperscript{S12}} & \multicolumn{3}{c}{simulations}\\
       \hline
& L & 2R & L/2R & R$_{z}$ & R$_{x(y)}$ & R$_{z}$/R$_{x(y)}$ \\ 
              \hline
       & \multicolumn{3}{c}{CS-356} & \multicolumn{3}{c}{CS-L} \\
              \hline
   core    & 330$\pm$12 & 75$\pm$8 & 4.4$\pm$0.1 & 14.1$\pm$0.1 & 3.2$\pm$0.1 & 4.4$\pm$0.14\\
microgel  & 627$\pm$69 & 372$\pm$41 & 1.7$\pm$0.4 & 31.4$\pm$0.5 & 16.9$\pm$0.3 & 1.8 $\pm$ 0.04\\
\hline
       & \multicolumn{3}{c}{CS-254} & \multicolumn{3}{c}{CS-M}\\
              \hline
   core    & 330$\pm$12 & 75$\pm$8 & 4.4$\pm$0.1 & 14.1$\pm$0.1 & 3.2$\pm$0.1 & 4.4 $\pm$ 0.14\\
microgel  & 569$\pm$23 & 316$\pm$13 & 1.8$\pm$0.1 & 24.7$\pm$0.3 & 13.8$\pm$0.2 & 1.78$\pm$0.03\\
              \hline
        & \multicolumn{3}{c}{CS-165} & \multicolumn{3}{c}{CS-S}\\
              \hline
   core    & -&-&-& 14.1$\pm$0.1 & 3.2$\pm$0.1 & 4.4$\pm$0.14\\
microgel  &- &-&-& 21.8$\pm$0.6 & 9.6$\pm$0.2 & 2.27$\pm$0.07\\
       \hline
     \end{tabular}
     \label{tab1aSim}
     \end{table}

\begin{table}[H]
    \centering
    \caption{Gyration radii of the microgels in bulk at T=20~°C.}
    \begin{tabular}{cccc}
    \hline
Gel   &  R$_{z}$ & R$_{xy}$ &  $R_{z}/R_{xy}$ \\
\hline
  CS-L	&  14.76$\pm$0.15&   7.50$\pm$0.11&	 1.97$\pm$0.02\\
 CS-M	&  11.06$\pm$0.05&	5.96$\pm$0.09& 1.86$\pm$0.02 \\
  CS-S & 9.47$\pm$0.10&  4.06$\pm$0.07 & 2.40$\pm$0.02\\
       \hline
     \end{tabular}
     \label{tab1bSim}
     \end{table}

\begin{table}[H]
     \centering
    \caption{Types and amount of beads in microgels. The beads forming polymer microgel shell (including crosslinkers) and the solid core are denoted as M and P respectively.}
     \begin{tabular}{cccc}
       \hline
       & Amount of M beads & Amount of P beads & \% of cross-linkers\\
\hline
CS-L	& 30005&	1939&	4.873\\
CS-M	&15010&	1939&	4.677\\
CS-S &	6005&	1939&	4.763\\
       \hline
     \end{tabular}
     \label{tab2Sim}
     \end{table}

\begin{table}[H]
     \centering
    \caption{Types and amount of bonds in microgels. Type 1 - bonds between MM and MP beads. Type 2 – bonds between PP beads. The beads forming polymer microgel shell (including cross-linkers) and the solid core are denoted as M and P respectively.}
     \begin{tabular}{ccc}
       \hline
& Type 1 & Type 2\\
\hline
CS-L &	31063+144 &	3431\\
CS-M &	15432+144	& 3431\\
CS-S &	6071+144&	3431\\
       \hline
     \end{tabular}
     \label{tab3Sim}
     \end{table}

\begin{table}[H]
     \centering
    \caption{Types and amount of angles in microgels. Type 1 - the angle between PPP beads. P is the bead forming the solid core.}
     \begin{tabular}{cc}
       \hline
& Type 1\\
\hline
Microgel M & 4634	\\
Microgel S&	4634\\
Microgel XS& 4634\\
       \hline
     \end{tabular}
     \label{tab4Sim}
     \end{table}

\begin{table}[H]
     \centering
    \caption{Hansen solubility parameters, T ~ 20°C*.}
     \begin{tabular}{cccc}
       \hline
Compound & $\delta_{i}^{d}$ & $\delta_{i}^{p}$ & $\delta_{i}^{h}$ \\
\hline
Water*	&15.5&	16.0&	42.3\\
Decane**	&16.96	&1.27	&0.29\\
NIPAm***	&19.2	&1.9	&12.3\\
Silicon Oxide****&	19.80&	27.26	&16.56\\
       \hline
     \end{tabular}\\
     \text{$\delta$ are represented in units of (J/cm$^{3}$)$^{0.5}$.}
    \begin{footnotesize} \text{* from reference\textsuperscript{S2} ; ** from reference\textsuperscript{S7} ; *** from reference\textsuperscript{S5} ; **** calculated from reference\textsuperscript{S4}}\end{footnotesize}
     \label{tab5Sim}
     \end{table}

\begin{table}[H]
     \centering
    \caption{DPD interaction parameters used in the present work, T = 20°C.*}
     \begin{tabular}{ccccc}
       \hline
       $a_{ij}$&M&P&W& O\\
\hline
M	&25	&46.68492	&25.6	&30.38943\\
P	&	&25&	53.24808&	56.81993\\
W	&	& &	25&	90.12635\\
O	&	&	& &	25\\
       \hline
     \end{tabular}
     \text{*$a_{ij}$ for W, D, M, P is represented in units of $k_{B}T/r_{c}$.}
     \label{tab6Sim}
     \end{table}

\begin{table}[H]
     \centering
    \caption{Parameters of the system used in the present work.}
     \begin{tabular}{ccc}
\hline
Total number density&	$\rho =N/V$	&3\\
Time step	&$\Delta t$	&0.01\\
Noise amplitude	&$\sigma$&	3\\
       \hline
      Bonds potential, $U_{ij}^{bonds}$&&\\
      $r_{eq}$ & &0.7 \\
      $k_{bond~MM}$ = $k_{bond~MP}$ & &10 \\
       $k_{bond~PP}$ & &100 \\
       \hline
      Angle potential, $U_{ij}^{angles}$&&\\
     $\Theta$ & &109.47 \\
$k_{bend~PPP}$ & &100 \\
  \hline
      \end{tabular}
     \label{tab7Sim}
     \end{table}

\textbf{Calculations}\\
In our study, we calculated various shape descriptors derived from the gyration tensor and LSQR approximation of surface, which describe the geometry of the microgel at different temperatures.\\
\underline{Gyration tensor} We deﬁne the gyration tensor as

\begin{equation}
    \boldsymbol{S} = \frac{1}{N} \sum\nolimits_{i = 1}^{N} \boldsymbol{s}_{i}\boldsymbol{s}_{i}^{T} = \overline{\boldsymbol{ss}^{T}} = \left[ \begin{array}{rrr}
\overline{x^{2}} & \overline{xy} & \overline{xz}\\ 
\overline{yx} & \overline{y^{2}} & \overline{yz} \\
\overline{zx} & \overline{zy} & \overline{z^{2}} \\ 
\end{array}\right],
    	\label{simeq9}
\end{equation}
where $\boldsymbol{s}_{i} = \left[ \begin{array}{r}
x_{i} \\ 
y_{i} \\
z_{i}\\ 
\end{array}\right] $ is the position vector of each bead, which is considered with respect to the center of mass of the microgel $\sum\nolimits_{i = 1}^{N} \boldsymbol{s}_{i} = 0$ , and the overbars denote an average over all beads $N$ in the microgels. To get a complete characterization of the elliptic microgels, we computed the eigenvalues of the gyration tensor $\lambda_{x} = \overline{X^{2}}$, $\lambda_{y} = \overline{Y^{2}}$, $\lambda_{z} = \overline{Z^{2}}$ as well as eigenvectors $\overline{T_{1}}$,$\overline{T_{2}}$,$\overline{T_{3}}$. The gyration tensor is symmetric thus, the Cartesian coordinate system can be found in which it is diagonal. Transformation to the principal axis system diagonalizes $S$, and we choose that principal axis system in which

\begin{equation}
    \boldsymbol{S} = diag(\lambda_{x},\lambda_{y},\lambda_{z}),
    	\label{simeq10}
\end{equation}
where we assume that the eigenvalues of $\boldsymbol{S}$ are sorted in ascending order, i.e., $\lambda_{x}\leq\lambda_{y}\leq\lambda_{z}$ and $x$, $y$ and $z$ are the new coordinate axes. These eigenvalues are called the principal moments of the gyration tensor. The first invariant of S gives the squared radius of gyration,

\begin{equation}
    tr(\boldsymbol{S}) \equiv I_{1} = \lambda_{x} + \lambda_{y} + \lambda_{z} = R_{g}^{2},
    	\label{simeq11}
\end{equation}
a measure of the average size of the particular conformation. Microgel consisting of beads of equal mass have by definition a moment of inertia tensor (concerning the center of gravity), $\boldsymbol{I}$, which is diagonal in the same frame of reference as $\boldsymbol{S}$.\\
By analogy with Theodorou et al.\textsuperscript{S13} we define the shape anisotropy of microgel in a particular conformation as the traceless deviatory part of $\boldsymbol{S}$

\begin{equation}
    \hat{\boldsymbol{S}} = \boldsymbol{S} - \frac{1}{3}tr(\boldsymbol{S})\boldsymbol{E},
    	\label{simeq12}
\end{equation}
 $tr(\hat{\boldsymbol{S}}) = 0$ where $\boldsymbol{E}$ is the unit tensor. In the principal axis system $\hat{\boldsymbol{S}}$, could be presented by a sum of two terms (in a way similar to molecular polarizability tensor\textsuperscript{S14})
 
 \begin{equation}
    \hat{\boldsymbol{S}} = B~\text{diag}(2/3, -1/3, - 1/3) + C~\text{diag}(0, 1/2, -1/2),
    	\label{simeq13}
\end{equation}
 where we deﬁne quantities such as the asphericity, $B$, and the acylindricity, $C$, of a Janus-like microgel, derived from the values of the principal moments

 \begin{equation}
   B = \lambda_{z} - \frac{1}{2}(\lambda_{x} + \lambda_{y}),
    	\label{simeq14}
\end{equation}

 \begin{equation}
   C = \lambda_{y} + \lambda_{x}),
    	\label{simeq15}
\end{equation}
which are always not negative. When the bead distribution within the microgel is spherically symmetric or has a cubic, a tetrahedral, or higher symmetry, then the three principal moments are equal, $\lambda_{x}$ = $\lambda_{y}$ = $\lambda_{z}$ and $b$ = 0. The acylindricity is zero only when the two principal moments are equal, $\lambda_{x}$ = $\lambda_{y}$. This zero condition is met when the distribution of particles is cylindrically symmetric and whenever the particle distribution is symmetric with respect to the two coordinate axes. In our study, we use dimensionless relative quantities which were taken concerning $I_{1}$

 \begin{equation}
  b = \frac{1}{TrS}B,\hspace{2cm} 0 \leq b \leq 1
    	\label{simeq16}
\end{equation}

 \begin{equation}
  c = \frac{1}{TrS}C,\hspace{2cm} 0 \leq c \leq 1.
    	\label{simeq17}
\end{equation}
An overall measure of shape anisotropy is the quantity of $tr(\hat{\boldsymbol{S}}\hat{\boldsymbol{S}}$),

 \begin{equation}
 tr(\hat{\boldsymbol{S}}\hat{\boldsymbol{S}}) = \frac{2}{3}B^{2} + \frac{1}{2}C^{2} = \frac{2}{3}I^{2}_{1} - 2I_{2},
    	\label{simeq18}
\end{equation}
where $I_{2}=\lambda_{x} \lambda_{y}+\lambda_{x} \lambda_{z}+\lambda_{y} \lambda_{z}$ is the second invariant of $S$. The dimensionless relative shape anisotropy $\kappa^{2}$ is defined as

 \begin{equation}
 \kappa^{2} = \frac{3}{2} tr\frac{(\hat{\boldsymbol{S}}\hat{\boldsymbol{S}})}{tr(\boldsymbol{S})^{2}} = 1 - \frac{3I_{2}}{I_{1}^{2}} = 1 - 3\frac{\lambda_{x}\lambda_{y}+\lambda_{x}\lambda_{z}+\lambda_{y}\lambda_{z}}{(\lambda_{x} +\lambda_{y}+\lambda_{z})^{2}}.
    	\label{simeq19}
\end{equation}
It reflects both the symmetry and dimensionality of a microgel. This parameter is also limited between the values of 0 and 1. It reaches 1 when all beads are located in a straight line and drop to zero for highly symmetric conformations.\\
\underline{The measure of orientational order} Upon compression, a monolayer of anisotropic particles becomes ordered. The ellipsoidal shape, the softness of the shell, and the compression within the trough reveal the different orientations of the microgels at the liquid-liquid interface. The description of the orientation order in the system is based on the analysis of the directions of the long axis of the solid core in the microgels. By analogy with the description of the liquid crystal phase behavior of solid anisotropic particles, we introduce the unit vector $\overline{n}$ called the director. In general, the direction of $\overline{n}$ is the average orientation direction of all particles in the system. The measure of the amount of order is the scalar orientational order parameter, defined as 

\begin{equation}
  S_{3D} = \frac{1}{2}\langle 3cos^{2}\Theta - 1 \rangle, \hspace{0.5cm} \text{in 3D}
    	\label{simeq20}
\end{equation}

\begin{equation}
  S_{2D} = \langle2cos^{2}\Theta - 1\rangle, \hspace{0.5cm} \text{in 2D}
    	\label{simeq21}
\end{equation}
where $\Theta$ is the angle between the long axes of the ellipsoid and the director $\overline{n}$. The scalar order parameter $S_{3D}$ can take any value between -1/2 to 1. If the long axis of all ellipsoids is aligning exactly with the director, so $\Theta$=0 for all molecules, $S_{3D}$ = 1. When all normal lie in the plane perpendicular to the director, but randomly oriented in that plane, then $\langle cos^{2}\Theta\rangle$ = 0, and $S_{3D}$ = -1/2. In the isotropic phase the molecules are randomly oriented, so $\langle cos^{2}\Theta\rangle$ =1/3, and S = 0. Such description is valid only if we have to deal with the uniaxial arrangement of uniaxial molecules. Moreover in the isotropic phase  $S_{3D}$  $\sim$ 1/$\sqrt{N}$, where $N$ is the number of particles.\\
In 2D $S_{2D}$ = 1 can take any value between 0 and 1. $S_{2D}$ = 1 if so $\Theta$=0 and $S_{2D}$= 0 for a completely disordered system. In our system, we choose the director that coincides in the direction of compression.\\

\begin{figure}[ht]
\centering
    \includegraphics[width=0.5\linewidth, trim={0cm 0cm 0cm 0cm},clip]{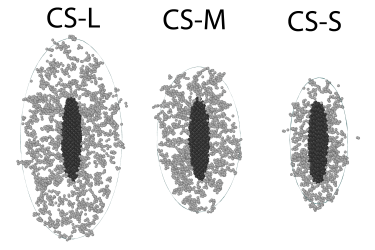}	
	\caption{Equilibrium structure of the swollen microgels with different thickness of the shell. The cross-sections of the thickness of $d = 2r_{c}$.  Polymeric shell depicted in grey, solid nanoparticle – in black, water beads are not shown.}
	\label{FigS1Sim}
\end{figure}

\textbf{Microgel characterization in bulk}\\
The cross-sections of the equilibrated microgels of the thickness of $d = 2r_{c}$ taken through the center of mass are shown in Figure \ref{FigS1Sim}. The sizes of the solid nanoparticle are  3.2$\pm$0.1$r_{c}$ $\times$ 3.2$\pm$0.1$r_{c}$ $\times$14.1$\pm$0.1$r_{c}$.

\begin{figure}[ht]
\centering
    \includegraphics[width=1\linewidth, trim={0cm 0cm 0cm 0cm},clip]{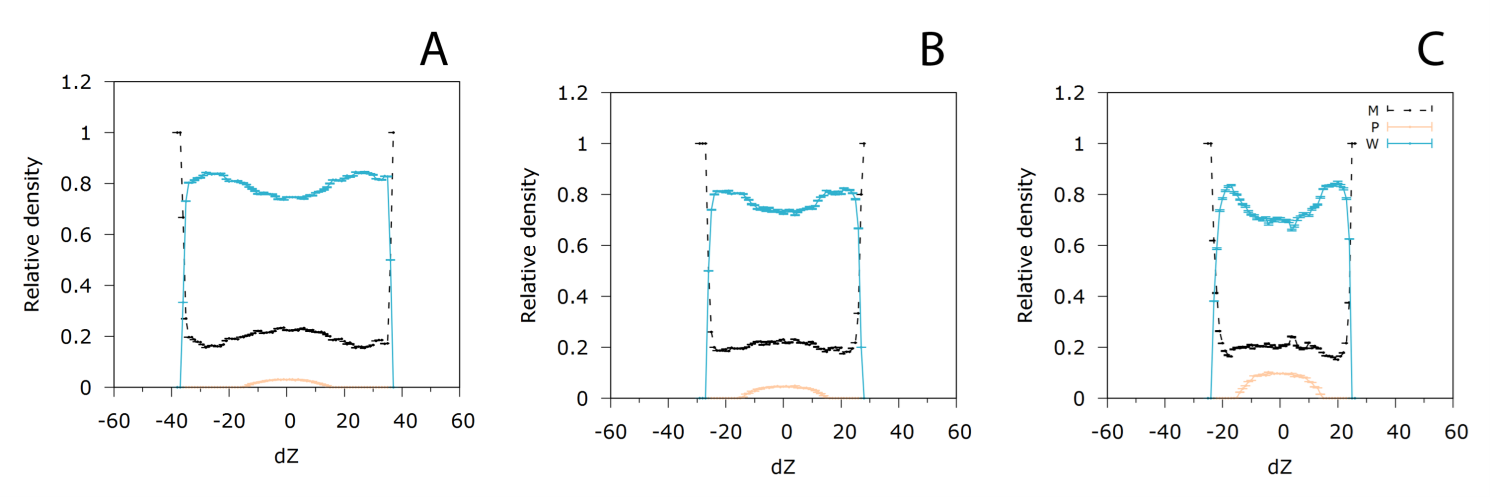}	
	\caption{Concentration profiles along the large axis of the ellipse. Black, yellow and blue curves correspond to the distribution of polymer, M, solid particle, P, and water beads, W, respectively. A, B and C corresponds to the microgels with different shell thickness: CS-L, CS-M and CS-S.}
	\label{FigS2Sim}
\end{figure}

With the selected values of the interaction parameters, which corresponds to the $T$=20°C, the microgels are in a swollen state (Figure \ref{FigS1Sim}, \ref{FigS2Sim}). The microgels contain approximately 80\% of the solvent (Figure \ref{FigS2Sim}).

In Table \ref{tab1aSim} and \ref{tab1bSim}, we present the comparison of the sizes of the major $L_{SANS}$ and minor axes 2$R_{SANS}$ of the elliptical silica core (and microgels) in a swollen state obtained by the SANS measurements and in the simulations. In the simulations, we fitted the surface of the microgels using LSQR approximation and estimate its radii. Also to make a more accurate comparison between the experiment and simulations, we calculated the relative sizes of the microgel normalized to the small principal axis of the solid nanoparticle, Table \ref{tab1bSim}. We choose such normalization since the sizes of the nanoparticle are independent of the system. We found an excellent correlation of the aspect ratio $R_{z}/R_{x(y)}$ both for the solid nanoparticles and for the microgels. Moreover we note the good agreement between of $L / R_{core} (R / R_{core})$ and $R_{z}$ / $R_{x(y)_{core}} (R_{x(y)} / R_{x(y)_{core}})$ respectively. Thus, we can confidently state that we are considering microgels that are identical in their characteristics in water at 20°C.\\
\textbf{Effect of the box and influence of the method of preparation of compressed systems}\\
To check whether the shape of the box influences the ordering of the microgels, we performed the additional simulations of the system in the simulation cells with a square base (isotropic shape of the box). We took 16 CS-M gels (the directions of the main axis of the microgels were random) and placed it to the four simulation cells of the same $L{z}$ but different interfacial area. Namely:\\
A) $L_{x}~\times~L_{y}=130~\times~130$, which relates to $80~\times~210$ (highly compressed system)\\
B) $L_{x}~\times~L_{y}=159~\times~159$, which relates to $120~\times~210$\\
C) $L_{x}~\times~L_{y}=171~\times~171$, which relates to $140~\times~210$\\
D) $L_{x}~\times~L_{y}=194~\times~194$, which relates to $180~\times~210$ (weakly compressed system)\\ 
Such procedure allowed us to prepare the systems with four different compression degrees $1/Ap = 9.5\cdot10^{-4}$(1); $6.3\cdot10^{-4}$(2); $5.4\cdot10^{-4}$ (3) and $4.7\cdot10^{-4}$ (4). The choice of such boxes is predicated on the fact that at corresponding $1/Ap$ values we observed a transition from the disordered to ordered states. Each simulation box was equilibrated independently of the others.\\
In order to provide a quantitative measure of the degree of orientational order in the layer plane, we used the order parameter of 2D nematics calculated from the largest positive eigenvalue of the order parameter tensor Q:\\
\begin{equation}
\boldsymbol{Q}_{\alpha\beta} = 1/N \displaystyle\sum_{i=1}^{N} (2e_{i\alpha}e_{i\beta}-\delta_{\alpha\beta}),
    	\label{simeq2}
\end{equation}
where $e_{i}$ is the normalized vector representing the $i^{th}$ molecular long axis $e_{ix}^2+e_{iy}^2=1$ and the subscripts $\alpha$ and $\beta$ are the coordinates x and y, respectively.\\
We observe the same trends for the evolution of the 2D order parameter Figure \ref{FigS14Sim}A (black dots). We confirm that starting from the end of phase II, a transition from a disordered state at high available areas, to a more ordered state occurs independently of the configuration of the simulation box as well as the method of preparation of the compressed state of the system. We also confirm the presence of the peak of the order parameter. Moreover, we also observe the decrease of the order parameter with further compression. Nevertheless, we have to admit that there is a difference in the direction of the director. It relates to the method of preparation of the system at different compression degrees from scratch. There are no contradictions there due to the periodic boundary conditions in xy direction used in simulations. Second, there is a difference in the magnitude of the peak of the order parameter. Using the method which reproduces the experimental technique allow us to perform smooth sequential change of the side of the cell and equilibrating the system each compression cycle which helps to reach more precise order parameter values.\\

\begin{figure}[ht]
\centering
    \includegraphics[width=0.9\linewidth, trim={0cm 0cm 0cm 0cm},clip]{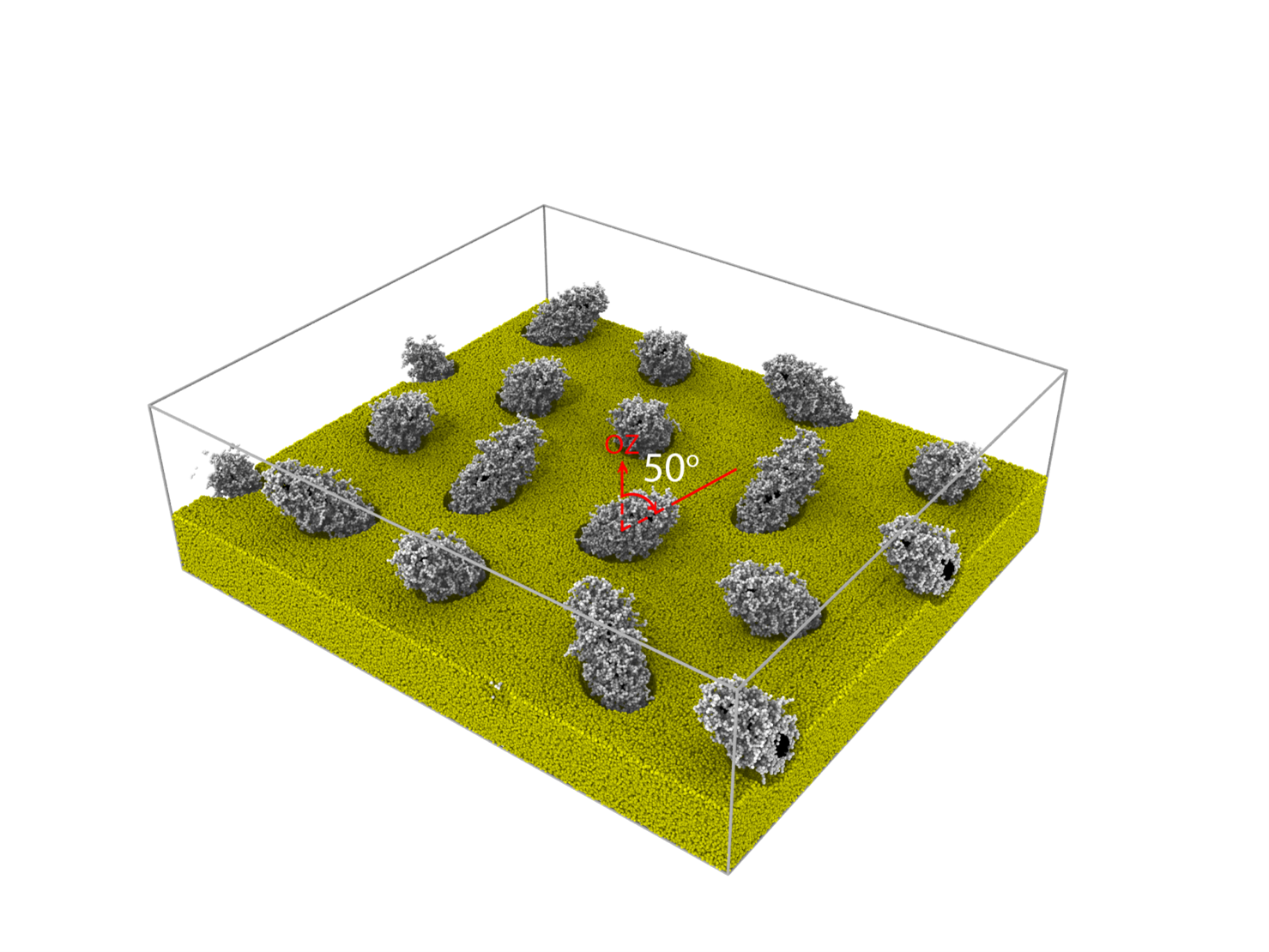}	
	\caption{3D views of the initial positioning of the gels at liquid/liquid interface. The grey, black and yellow beads correspond to the microgel shell (M), solid nanoparticle (P), and oil (O), respectively. Water beads are not shown. Each of the microgels is tilted with respect to the OZ axis to the angle of 50° and that is rotated to the random angle around the OZ axis.}
	\label{FigS3Sim}
\end{figure}

\begin{figure}[ht]
\centering
    \includegraphics[width=0.75\linewidth, trim={0cm 0cm 0cm 0cm},clip]{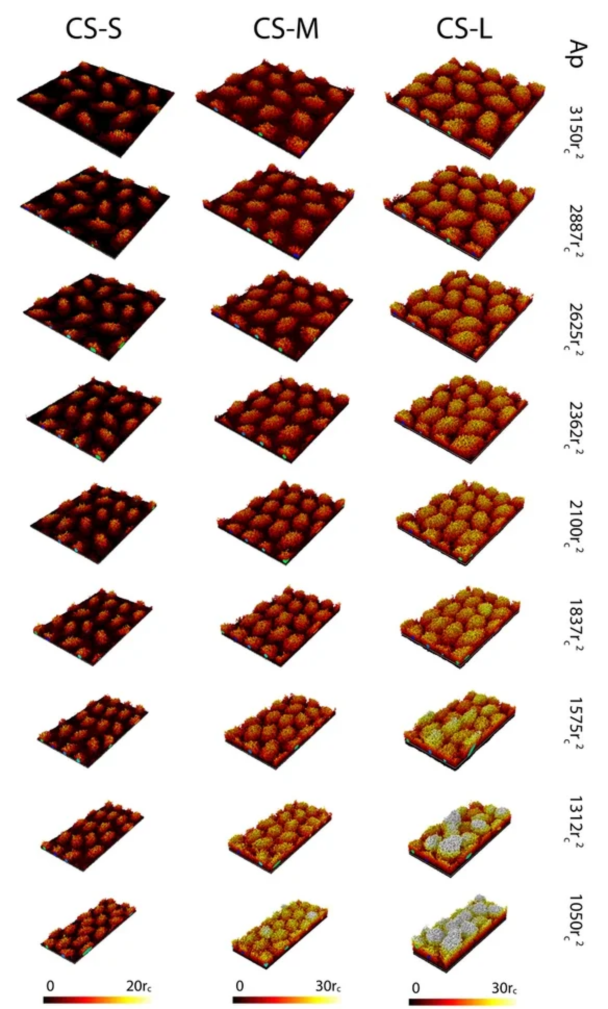}	
	\caption{3D height maps (as seen from the water phase) of simulation snapshots of adsorbed microgels having different shell thickness: CS-L, CS-M and CS-S. The different rows correspond to the different compression degrees (surface coverage). The microgels are swollen in water and collapsed in oil.}
	\label{FigS4Sim}
\end{figure}

\begin{figure}[ht]
\centering
    \includegraphics[width=0.5\linewidth, trim={0cm 0cm 0cm 0cm},clip]{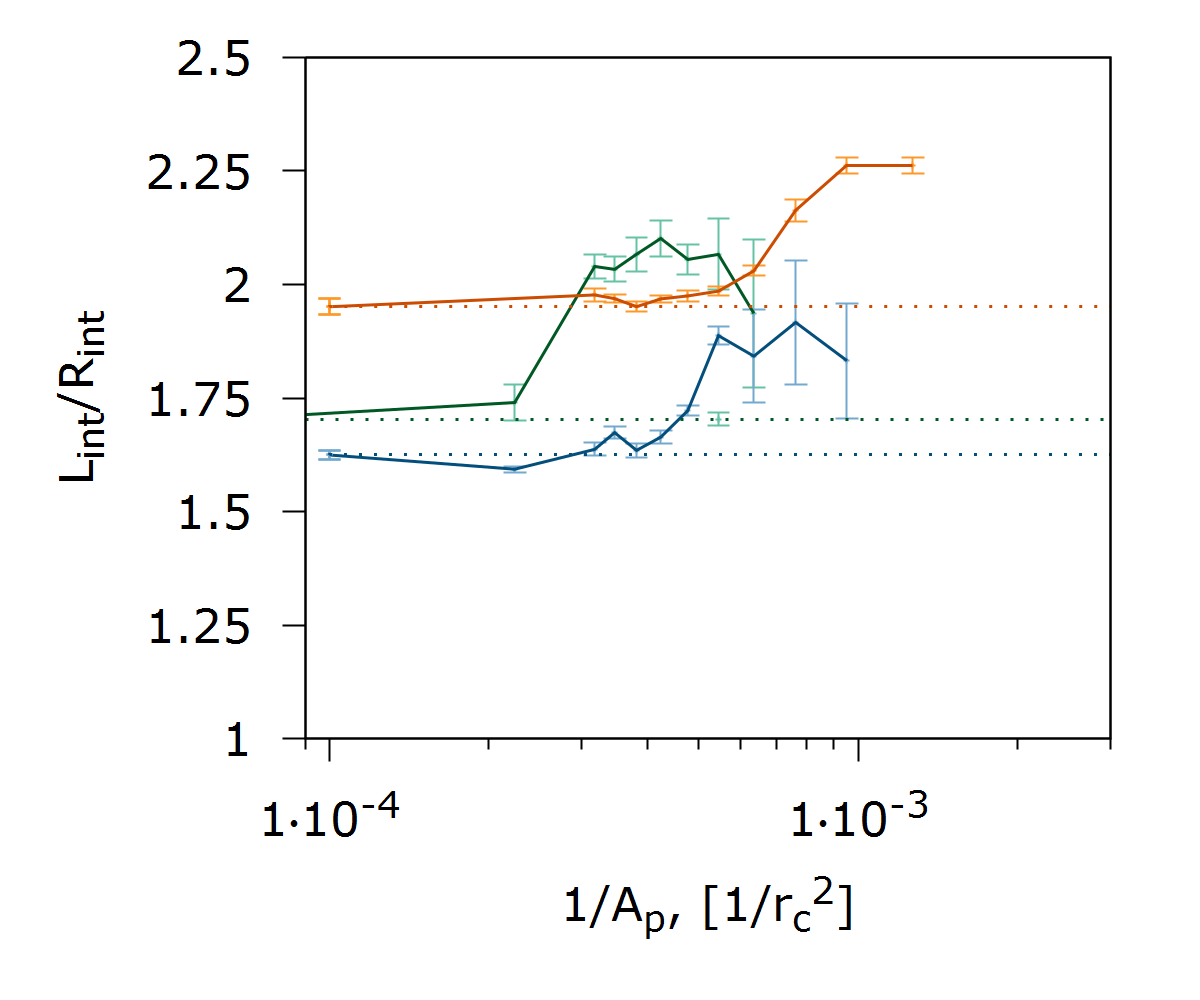}	
	\caption{Relative anisotropy of the microgels $L_{int}/R_{int}$ as a function of inverse area per particle of the microgel-covered oil-water interface. Solid and dashed lines correspond to the mean values of the long and short radii of the spreading area of the microgel.  The curves of the different colors correspond to the microgels of different shell thicknesses: CS-L (green), CS-M (blue) and CS-S (orange).}
	\label{FigS10Sim}
\end{figure}

\begin{figure}[ht]
\centering
    \includegraphics[width=1\linewidth, trim={0cm 0cm 0cm 0cm},clip]{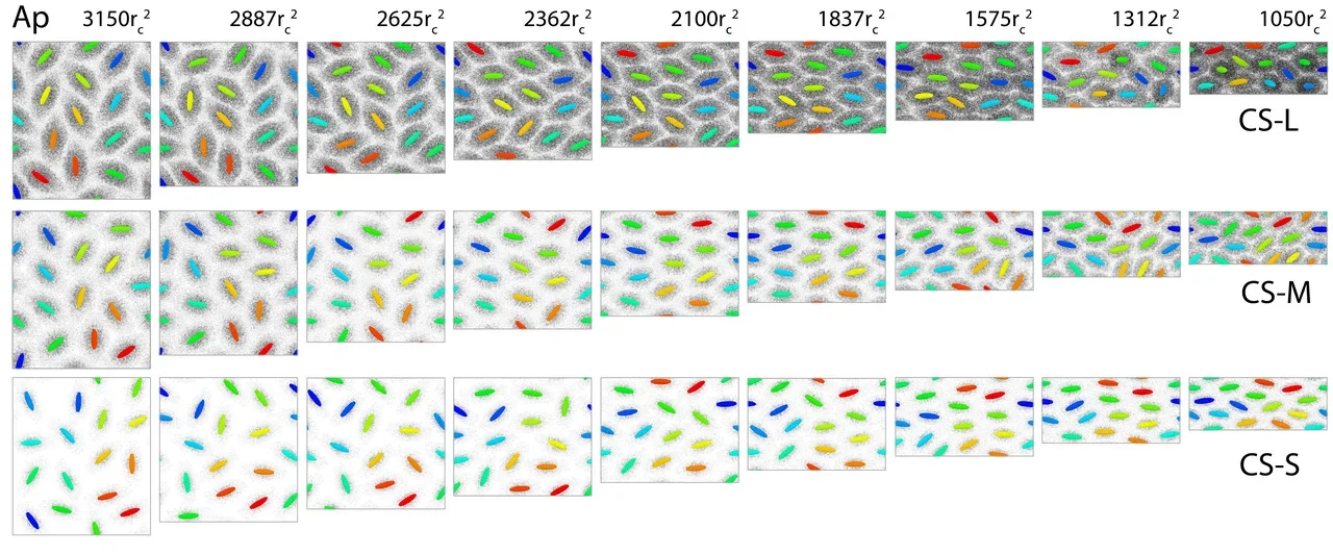}	
	\caption{Bottom view (as seen from the water phase) of simulation snapshots of adsorbed microgels having different shell thickness: CS-L, CS-M and CS-S. The different columns have different compression degrees (surface coverage). Each solid particle is colored in a different color. The black dots represent the polymer shell.}
	\label{FigS11Sim}
\end{figure}

\begin{figure}[ht]
\centering
    \includegraphics[width=0.65\linewidth, trim={0cm 0cm 0cm 0cm},clip]{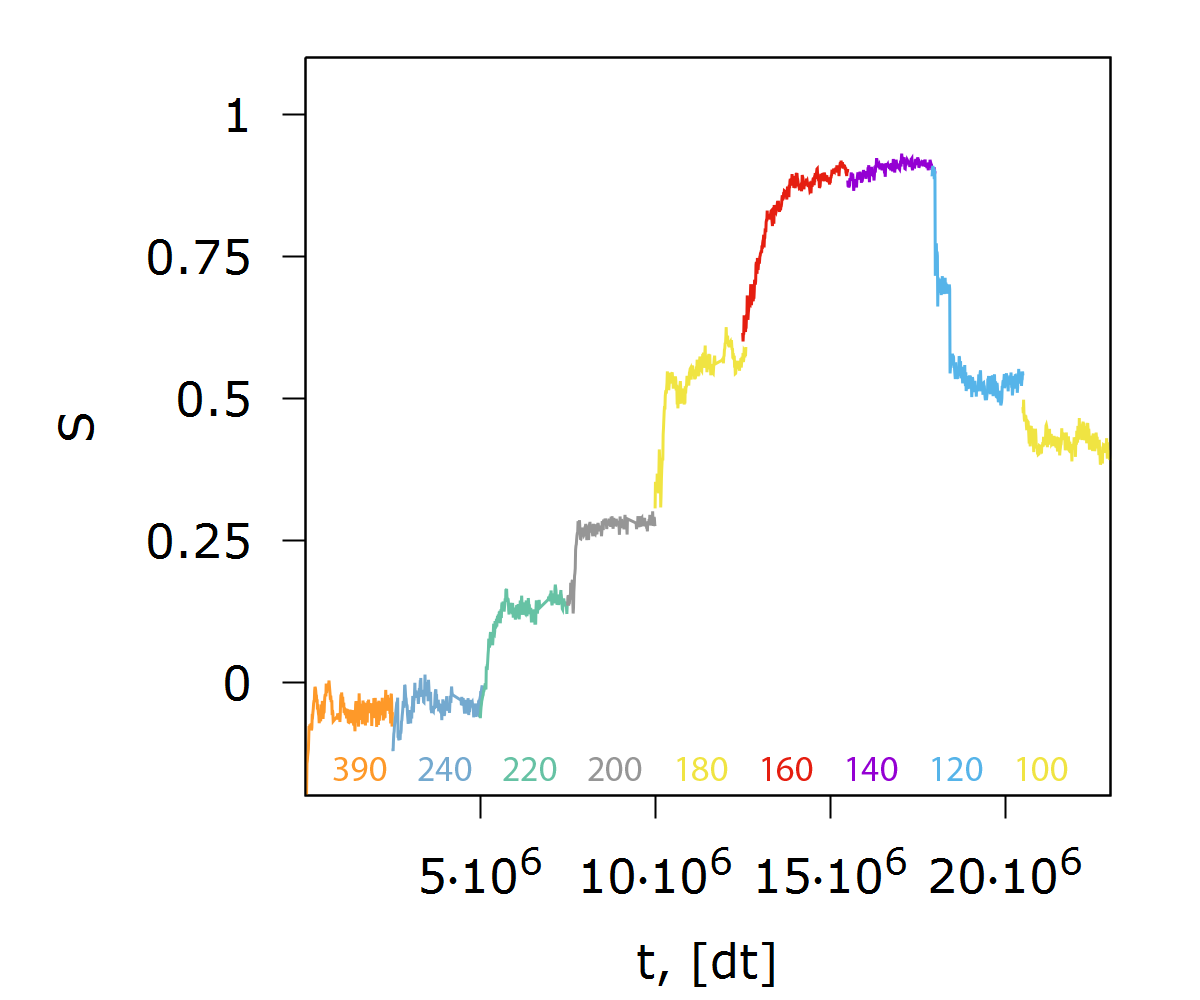}	
	\caption{Fluctuation curves of order parameter in the equilibration run for CS-M microgels. Different color corresponds to the different degree of compression ($L_{x}$ = 390, 240, 220, … 100).}
	\label{FigS13Sim}
\end{figure}

\begin{figure}[ht]
\centering
    \includegraphics[width=0.8\linewidth, trim={0cm 0cm 0cm 0cm},clip]{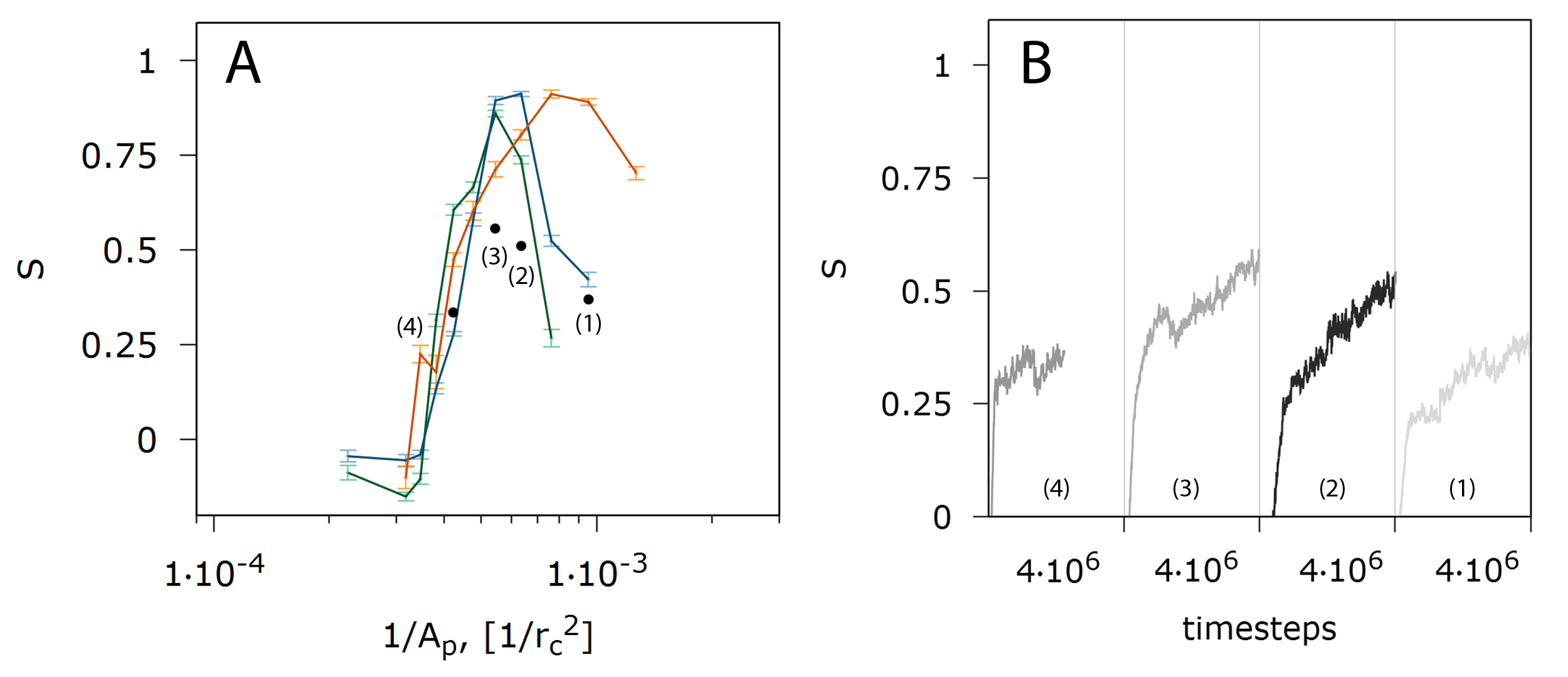}	
	\caption{A. Evolution of the 2D order parameter for the solid cores as a function of inverse area per particle of the microgel-covered oil-water interface. The curves of the different colors correspond to the microgels of different shell thicknesses: CS-L (green), CS-M (blue) and CS-S (orange). Black dotes corresponds to the order parameter for CS-M microgels which was simulated in the box with a square base $L_x=L_y$. B. Fluctuation curves of the 2D order parameter for CS-M microgels. Different color corresponds to the different degree of compression $1/Ap = 9.5\cdot10^{-4}$(1); $6.3\cdot10^{-4}$ (2); $5.4\cdot10^{-4}$ (3) and $4.7\cdot10^{-4}$ which are related to the box with $L_x~\times~L_y=130~\times~130$; $159~\times~159$; $171~\times~171$ and $194~\times~194$ respectively.}
	\label{FigS14Sim}
\end{figure}

\begin{doublespace}
\section{References}
(S1) Groot, R. D.; Warren, P. B. Dissipative Particle Dynamics: Bridging the Gap between Atomistic and Mesoscopic Simulation. \textit{J. Chem. Phys.} \textbf{1997}, \textit{107}, 4423–4435.\\
\\
(S2) Hansen, C. M. \textit{Hansen solubility parameters: a user’s handbook}; CRC press, 2007.\\
\\
(S3) Lindvig, T.; Michelsen, M. L.; Kontogeorgis, G. M. A Flory-Huggins Model Based on the Hansen Solubility Parameters. \textit{Fluid Phase Equilib.} \textbf{2002}, \textit{203}, 247–260.\\
\\
(S4) Faasen, D. P.; Jarray, A.; Zandvliet, H. J. W.; Kooij, E. S.; Kwiecinski, W. Hansen Solubility Parameters Obtained via Molecular Dynamics Simulations as a Route to Predict Siloxane Surfactant Adsorption. \textit{J. Colloid Interface Sci.} \textbf{2020}, \textit{575}, 326–336.\\
\\
(S5) Van Krevelen, D.; Te Nijenhuis, K. In \textit{Properties of Polymers (Fourth Edition)}, fourth edition ed.; Van Krevelen, D., Te Nijenhuis, K., Eds.; Elsevier: Amsterdam, \textbf{2009}; pp 3–5.\\
\\
(S6) Groot, R. D.; Rabone, K. L. Mesoscopic Simulation of Cell Membrane Damage, Morphology Change and Rupture by Nonionic Surfactants. \textit{Biophys. J.} \textbf{2001}, \textit{81}, 725–736.\\
\\
(S7) Alasiri, H.; Chapman, W. G. Dissipative Particle Dynamics (DPD) Study of the Interfacial Tension for Alkane/Water Systems by Using COSMO-RS to Calculate Interaction Parameters. \textit{J. Mol. Liq.} \textbf{2017}, \textit{246}, 131–139.\\
\\
(S8) Anderson, R. L.; Bray, D. J.; Ferrante, A. S.; Noro, M. G.; Stott, I. P.; Warren, P. B. Dissipative Particle Dynamics: Systematic Parametrization UsingWater-Octanol Partition Coefficients. \textit{J. Chem. Phys.} \textbf{2017}, \textit{147}, 094503.\\
\\
(S9) Yong, X.; Kuksenok, O.; Matyjaszewski, K.; Balazs, A. C. Harnessing Interfacially-Active Nanorods to Regenerate Severed Polymer Gels. \textit{Nano Lett.} \textbf{2013}, \textit{13}, 6269–6374.\\
\\
(S10) Kleinschmidt, D.; Fernandes, M. S.; Mork, M.; Meyer, A. A.; Krischel, J.;Anakhov, M. V.; Gumerov, R. A.; Potemkin, I. I.; Rueping, M.; Pich, A. Enhanced Catalyst Performance through Compartmentalization Exemplified by Colloidal LProline Modified Microgel Catalysts. \textit{J. Colloid Interface Sci.} \textbf{2020}, \textit{559}, 76–87.\\
\\
(S11) Plimpton, S. Fast Parallel Algorithms for Short-Range Molecular Dynamics. \textit{J. Comput. Phys.} \textbf{1995}, \textit{117}, 1–19.\\
\\
(S12) Nickel, A. C.; Scotti, A.; Houston, J. E.; Ito, T.; Crassous, J.; Pedersen, J. S.; Richtering, W. Anisotropic Hollow Microgels That Can Adapt Their Size, Shape, and Softness. \textit{Nano Lett}. \textbf{2019}, \textit{19}, 8161–8170.\\
\\
(S13) Theodorou, D. N.; Suter, U. W. Shape of Unperturbed Linear Polymers: Polypropylene. \textit{Macromolecules} \textbf{1985}, \textit{18}, 1206–1214.\\
\\
(S14) Smith, R. P.; Mortensen, E. M. Bond and Molecular Polarizability Tensors. I. Mathematical Treatment of Bond Tensor Additivity. \textit{J. Chem. Phys.} \textbf{1960}, \textit{32}, 502–507.

\end{doublespace}